\documentclass[pra,a4paper,aps,twocolumn,showpacs,superscriptaddress,
groupedaddress]{revtex4-2}
\usepackage{ae}
\usepackage[T1]{fontenc}
\usepackage[ansinew]{inputenc}
\usepackage{amsmath}
\usepackage{amssymb}
\usepackage{graphicx}
\usepackage{tikz,ragged2e}
\usepackage[caption=false]{subfig}
\usepackage{color,mathtools}
\usepackage[colorlinks=true,linkcolor=red,citecolor=blue]{hyperref}
\DeclarePairedDelimiter\bra{\langle}{\rvert}
\DeclarePairedDelimiter\ket{\lvert}{\rangle}

\DeclarePairedDelimiterX\braket[2]{\langle}{\rangle}{#1 \delimsize\vert #2}
\newcommand\beq{\begin{equation}}
\newcommand\eeq{\end{equation}}
\newcommand\bea{\begin{eqnarray}}
\newcommand\eea{\end{eqnarray}}
\begin{document}
	\title{Disorder operators and magnetic vortices in SU(N) lattice gauge theory}
	\author{Manu Mathur} 
\email{manu@bose.res.in,manumathur14@gmail.com}
\email{Visiting Scientist: The Institute of Mathematical Sciences,}
\email{Chennai, Professor (Retd.), S. N. Bose National Centre for Basic}
\email{Sciences, Kolkata; Present address: A-6/8 BCHS, Baghajatin,}
\email{Kolkata 700094.}
\author{Atul Rathor$^1$}
\email{atulrathor@bose.res.in,atulrathor999@gmail.com}
\affiliation{$^1$S. N. Bose National Centre for Basic Sciences,\\ Block JD, Sector III, Salt Lake, Kolkata 700106, India}
\begin{abstract}

\noindent  We construct the most general disorder operator for  SU(N)  lattice 
gauge theory in $(2+1)$ dimension by using  exact  duality transformations. 
These disorder operators, defined on the plaquettes and characterized by ($\text{N}-1$) angles, are the creation \& annihilation  or  the  shift operators for the SU(N) magnetic vortices 
carrying $(\text{N}-1)$ types of magnetic fluxes.
They are dual to the SU(N)
Wilson loop order operators which, on the other hand, are the creation-annihilation or shift operators for the $(\text{N}-1)$ electric fluxes on their loops.   
The new order-disorder algebra involving SU(N) Wigner D matrices
is derived and discussed.  The  $Z_\text{N} (\in $  SU(\text{N})) 't Hooft operator is obtained as a special limit.  In this limit  we also recover the standard Wilson-'t Hooft 
order-disorder algebra. The partition function representation 
and the free energies of these SU(N) magnetic vortices
are discussed.
\end{abstract}
\maketitle 
\section{Introduction}
Disorder operators, introduced originally in 1971 by Kadanoff and Ceva in the context of the two-dimensional Ising model \cite{kada},  have been widely discussed  and found useful in the studies of phase structures of  spin models as well as  abelian and non-abelian  gauge theories \cite{kramers,wegner,man1,hooft,mack,frad,dis_nonabelian,guth,kogut51,tomb,magnetic_disorder,reinhardt}. They also play a pivotal role in differentiating the topological phases of matter \cite{frad}  and in the  boson-fermion transmutation through the `order $\otimes$ disorder' combinations \cite{jordanwigner}.  
It is generally known that the duality transformations in spin models and gauge theories naturally lead to these disorder operators as the fundamental operators  describing  the dual interactions. Under duality, the interacting and the non-interacting terms also interchange their roles leading to the inversion of the coupling constant in the dual interactions. 
The Kramers-Wannier duality in $(1+1)$ dimensional Ising spin model \cite{kramers} and the Wegner duality in $(2+1)$ dimensional $Z_2$ gauge theory are the simplest examples which illustrate the above facts \cite{wegner,frad,guth,kogut51}. 
In the $(1+1)$ dimensional Ising model the disorder operators are  simply the dual spin operators which  describe the dual interactions with inverse coupling. 
They also create $Z_2$ kinks 
which are responsible for disordering the ground state leading to the loss of magnetization 
above the Curie temperature.

In abelian and non-ablian gauge theories the disorder operators acquire additional meaning of the dual electric potentials as the duality transformations also interchange the roles of the electric and magnetic degrees of freedom \cite{frad,wegner,guth,kogut51}.  Again, the Wegner dualities in the simplest $Z_\text{2}$ Ising gauge theory  in $(2+1)$ as well as in $(3+1)$ dimensions clearly illustrate  this additional rich feature \cite{frad,wegner,guth,kogut51}. More explicitly, in $(2+1)$ dimension $Z_2$ lattice gauge theory the disorder operators  are the dual spin or dual $Z_2$ electric potential operators \cite{wegner,guth} which describe the interactions in the dual formulation with inverse coupling. Being conjugate to the $Z_2$ magnetic fields, they also  create $Z_2$ magnetic vortices. These vortices, in turn,  magnetically disorder the ground states in the confining  phase \cite{wegner} and are thus responsible for the confinement-deconfinement phase transition. 

In general, the  order (disorder) operators are related to the potentials (dual potentials) which are conjugate to electric (magnetic) fields respectively.  They can  therefore be interpreted as  the ``translation operators"  for the electric and magnetic fluxes respectively. Moreover, the order-disorder algebra is simply the canonical commutation relations between the dual conjugate operators, i.e, between the magnetic flux and the electric potential operators (see the relations (\ref{mdoi}), (\ref{wlo}) and  (\ref{oda1})). 
In SU(3) lattice gauge theory or QCD the color confinement can be viewed as a consequence 
of magnetically disordered  ground state leading to area law for the Wilson loops. 
Like the various cases discussed above, the magnetic disorder in QCD is produced 
by the magnetic vortices, which in tun are  created by  the SU(3) disorder operators 
leading to disordered ground state.  A systematic study of these disorder operators in SU(2), SU(3) and then SU(N)  lattice gauge theories, using exact duality transformations, is the subject of this work. 

In 1978 Mandelstam tried to construct the SU(N) disorder operator  in the continuum using the dual electric non abelian vector potentials \cite{man1, hooft}. 
In 1978 't Hooft emphasised the role of disorder operators in the context of quark confinement in SU(N) gauge theory \cite{hooft}. The 't Hooft disorder operator creates  magnetic fluxes which 
belong to the center $Z_{\text{N}}$ of the gauge group SU(N).  
They have been extensively studied in the past analytically as well as using Monte Carlo techniques in the weak coupling continuum limit \cite{mack,tomb,dis_nonabelian}.

It is known that the 't Hooft loop disorder operators are dual to the 
Wilson loop order operators in a limited sense \cite{tomb,dis_nonabelian}
as they create only the center or $Z_{\text{N}}$  magnetic fluxes. 
In this paper we construct the most general disorder operator for SU(N) lattice gauge theory in ($2+1$) dimensions by exploiting the exact duality transformations \cite{mansre,mmar}. 
These disorder operators $\Sigma_{[\vec \theta]}(p)$
are defined on plaquettes $p$ as: 
\bea
 \Sigma^\pm_{[\vec \theta]}
 (p)= \exp \;i \big ( \vec \theta(p) \cdot \vec {\mathcal E}_\pm(p)\big ).
 \label{mdoi} 
\eea
In (\ref{mdoi}), $\vec {\cal E}_\pm(p)$ are the  SU(N) ``electric scalar potentials" on the plaquette $p$. They  are related to the SU(N) electric fields through the exact duality transformations (\ref{edt}) 
in Sec. \ref{dts} (also see  Figure \ref{fig:dual_relation}). 
The SU(N) disorder operator $\Sigma^\pm_{[\vec \theta]}(p)$ in (\ref{mdoi}) is characterized by a set of $(\text{N}-1)$  angles  which are denoted by 
$[\vec \theta \,] \equiv (\theta_1(p),\theta_2(p),\cdots ,\theta_{\text{N}-1}(p))$ on each plaquette. 
In this work, like the Kramers-Wannier spin and Wegner gauge dualities discussed earlier, we show that the  exact SU(N) duality transformations naturally lead to $\Sigma^\pm_{[\theta]}(p)$ in (\ref{mdoi}). We further show that they are the creation \& annihilation operators for the SU(N) magnetic votices 
on the spatial plaquettes. 

The  Wilson loop order operators 
 ${\cal W}^{[\vec j]} ({\cal C})$, on the other hand, are defines as 
 a path-ordered product of the link holonomies along a directed loop ${\cal C}$:
 \bea 
 {\cal W}^{[\vec j]} ({\cal C}) = \prod_{l \in {\cal C}}
 U^{[\vec j]}(l).  ~~~~~~
 \label{wlo} 
 \eea  
In (\ref{wlo}), $U^{[\vec j]}(l)$  are the SU(N) link holonomies or the ``magnetic vector potentials" in a general $[\vec j]$ representation of SU(N). 
Note that the SU(N) order operator ${\cal W}^{[\vec j]} ({\cal C})$ is characterized by a set of $(\text{N}-1)$  integers on loop ${\cal C}$
and $[\vec j] \equiv (j_1,j_2,\cdots ,j_{\text{N}-1})$.
The representation index 
$[\vec j]$  denotes  the  (N-1) eigenvalues $(j_1,j_2,\cdots ,j_{\text{N}-1})$ 
of the (N-1) SU(N) Casimir operators. These Casimir operators (constructed purely out of the electric 
field operators)  acting on the SU(N) electric basis  measure the net electric fluxes on the 
loop states created by the loop operator Tr $({\cal W}^{[\vec j]}({\cal C)})$. 
In this work we also obtain the SU(N) order-disorder operators algebra:
\begin{align}
\begin{aligned}  
\Sigma_{[\vec \theta]}(p)\;& 
\left({\cal W}^{[\vec j]}({\cal C})\right)_{\alpha\beta} \;
\Sigma^{-1}_{[\vec \theta]}(p) \\
~~~~&=\begin{cases}&\!\!\!\!\left(D^{[\vec j]}(\vec \theta)
~{\cal W}^{[\vec j]}({\cal C})\right)_{\alpha \beta},  ~~{\text{ if}} ~ p~ {\text{inside}}~ {\cal C}  \\
 &\!\! ~~ \left({\cal W}^{[\vec j]}({\cal C})\right)_{\alpha\beta}, ~~~~~~~~~~~~{\text{otherwise}}.
 \end{cases}
\end{aligned}\label{oda1}  
\end{align}  
In (\ref{oda1}), $D^{[\vec j]}(\vec \theta)$ denotes the SU(N) Wigner rotation matrix in the $[\vec j]$ representation. 
If the angles $[\vec \theta]$  correspond to the  centre element $z \in Z_\text{N}$ with  $z^\text{N}=1$, then  using \footnote{This result  follows from the SU(N) Young tableau in the $[\vec j]$ representation with total $L$ fundamental boxes. If each of them is rotated by the center element $z$ then we get $D^{[\vec j]}(z) = (z)^{L} \;{\cal I} = (z)^{\eta[\vec j]} \;{\cal I}$ as $z^\text{N}=1$. Here ${\cal I} $ is the identity matrix in the  $[\vec j]$ representation of SU(N).}  $D^{[\vec j]}(z) = (z)^{\eta[\vec j]}$, where $ \eta[\vec j] \left(=0,1,2\cdots, (\text{N}-1)\right)$ is the N-ality of the representation $[\vec j]$, we recover the standard 't Hooft-Wilson order-disorder algebra discussed  in  \cite{hooft}.  



The plan of the paper is as follows. In Sec. \ref{hdual} and Sec. \ref{dts} 
we summarise the Hamiltonian framework and the SU(N) duality transformations respectively.  These sections are only for setting up the notations
and to explain  the SU(N) duality relations which are then used directly in the  following sections. The details can be found in \cite{mmar,mansre}.  The SU(N) magnetic vortex creation and annihilation or equivalently SU(N) disorder operators are discussed in section \ref{sundo}.  In order to simplify the presentation, the SU(2), SU(3) and SU(N) disorder operators 
are discussed  one by one in the increasing order of difficulty  in sections \ref{su2disop}, \ref{su3disop} and \ref{sundisop} respectively. 
In the simplest SU(2) case, we construct the  magnetic basis in Section \ref{su2disop}-A using the SU(2) prepotential approach \cite{manpp}. 
In Section \ref{su2disop}-B we show that the SU(2) disorder operators act as SU(2) magnetic vortex creation-annihilation operators on the magnetic basis. The SU(2) order-disorder algebra is discussed in 
\ref{su2disop}-C. Some of the results in this section can also be found in \cite{mansre}. We then consider the SU(3) case in detail in \ref{su3disop}. As expected, there are many new SU(3) features which are absent in the simple SU(2) case.  In particular, we emphasize the importance of the  SU(3) prepotential operators  representation of the dual electric scalar potentials for constructing the SU(3) magnetic fields.  In Sec. \ref{su3disop} we directly generalize these SU(3) results to the SU(N) case. In Sec. \ref{ids}, we rewrite the SU(N) disorder operator in the original Kogut-Susskind formulation. We  show that they now become non-local operators and are attached with the invisible SU(N) Dirac strings. As expected,  these unphysical strings can be moved around by SU(N) gauge transformations without changing their  end points which specify the locations of the SU(N) gauge invariant magnetic vortices and anti-vortices. 
In Sec. \ref{pathrep} we  compute the path integral expression for the SU(N) vortex-free energy. This path integral representation should be useful for Monte Carlo simulations and 
to understand the role of these magnetic vortices and their condensation, if any, in the color confinement problem. It is expected that they will condense and disorder the vacuum state for any non-zero coupling constant.  

 The prepotential operators  create and  annihilate the SU(N) electric as well as the  magnetic fluxes \cite{manpp}. Therefore they provide  a common  platform to construct both the  electric and  magnetic bases in the physical loop Hilbert space of SU(N) lattice gauge theory.  In these two dual bases we show that the order and disorder operators have natural action of translating the electric and magnetic fluxes respectively. These SU(N) electric and magnetic bases and the action of the order and the disorder operators on them are discussed in detail in Appendix \ref{elb} and Appendix \ref{mlb} respectively. Appendix \ref{app:invisibility} shows that the SU(N) Dirac strings are unphysical.

 As mentioned earlier, we work in $(2+1)$ dimension. 
 The notations used are as follows: the lattice sites 
 are denoted by $(\vec n) \equiv (m,n)$ and the links by ${l} =(\vec n, \hat i)$ where 
 $\hat i =1,2$ denotes unit vectors in the two spatial directions. All the initial operators are vectors and assigned  to the links $l$.  All the dual operators are scalars and are defined on the plaquettes $(p)$ of the spatial two dimensional lattice. Many times we will suppress  the plaquette indices $(p)$ on the dual operators  to avoid clutter. 



 
\section{Hamiltonian Formulation}
\label{hdual}
In this section, we briefly discuss SU(N) Kogut-Susskind Hamiltonian lattice gauge theory
in (2+1) dimension.  The Hamiltonian of SU(N) lattice gauge theory is \cite{kogut51,kogsus}
\begin{equation}\label{ham}
H = \sum_{\vec n,\,\hat i} E^2(\vec n;\, \hat i)+ K \sum_{p} \text{Tr} \left(U_{p} + U^{\dagger}_{p}\right)   
\end{equation}
In equation (\ref{ham}), $E^2(\vec n;\,\hat i\,) \equiv \sum_{{\rm a}=1}^{\text{N}^2-1} \left(E_\pm^{\rm a}(\vec n;\,\hat i\,) \right)^2$ , 
 $U_{p} \equiv U(\vec n;\,\hat i\,)U(\vec n+\hat i;\,\vec j\,)U^{\dagger}(\vec n+\hat j;\,\hat i\,)U^{\dagger}(\vec n;\,\hat j\;)$,  and $K$ is a coupling constant. This is an electric field and magnetic vector potential description in which each link $(\vec n;\,\hat i\,)$ carries a SU(N) link flux operator  $U(\vec n;\,\hat i\,)$. We call $U(\vec n;\,\hat i\,)$ the link holonomy.  Their  left and  right link electric 
 fields $E^{\rm a}_\pm(\vec n;\,\hat i\,)$ 
rotate the link holonomies $U(\vec n;\,\hat i\,)$ from the left and right respectively or equivalently satisfy  the following commutation relations:
\begin{align}
\begin{aligned}\label{com_EU}
[E^{\mathrm a}_+(\vec n;\hat i\,),\;U(\vec n;\hat i\,)_{\alpha\beta}]=&\left( T^{\mathrm a}U(\vec n; \hat i\,) \right)_{\alpha\beta},\\
[E^{\mathrm a}_-(\vec n+\hat i;\hat i\,),\;U(\vec n;\hat i\,)_{\alpha\beta}]=& -\left( U(\vec n; \hat i\,) T^{\mathrm a}\right)_{\alpha\beta}
\end{aligned}
\end{align}
where $T^{\mathrm a}, \mathrm{a}=1,2,\cdots,\text{N}^2-1$ are the generators of fundamental representation of SU(N). 
These left and right electric fields are not independent and  are related by the link holonomy parallel transport
\begin{equation}
E_-(l) = -U^\dagger (l)\, E_+(l)\, U(l),
\label{lrrelxx} 
\end{equation}
In (\ref{lrrelxx})  $E_\pm(l) \equiv \sum_{\mathrm{a}=1}^{\text{N}^2-1}E^\mathrm{a}_\pm(l)T^\mathrm{a}$.
The commutation relations  (\ref{com_EU}) and Jacobi identity imply the electric fields  $E^{\mathrm a}_\pm(\vec n;\hat i)$ follow the SU(N) Lie algebra
\begin{align}
\begin{aligned}
[E^{\rm a}_+(\vec n;\hat i\,),\;E^{\rm b}_-(\vec n+\hat i;\hat i\,)] ~&= ~~~0, \\
[E^{\mathrm a}_\pm(\vec n;\hat i\,),\;E^{\mathrm b}_\pm(\vec n; \hat i)\,]~~~~&=~i\, f^{\mathrm{abc}}E^{\mathrm c}_\pm(\vec  n;\hat i\,).
\label{com_EE}
\end{aligned}
\end{align}
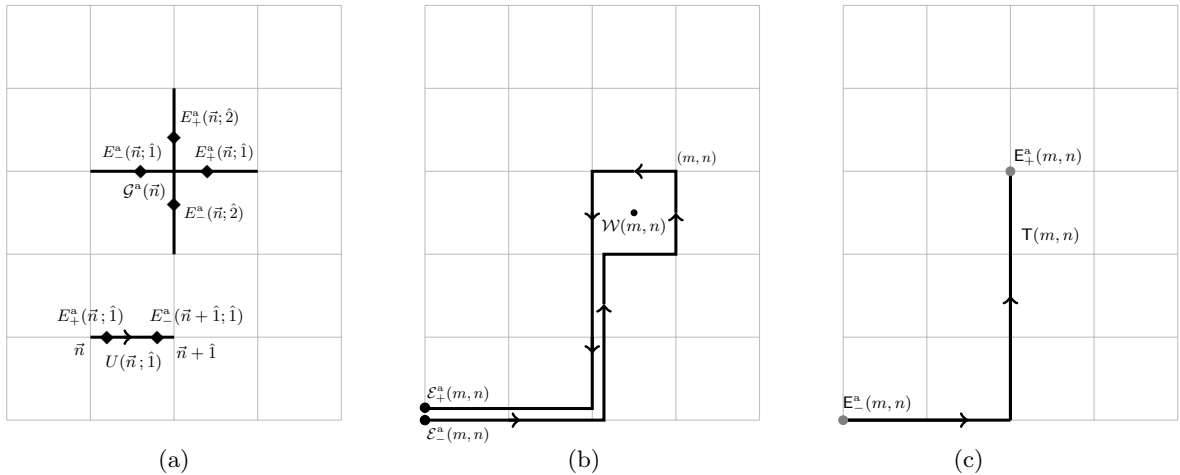
\begin{figure*}[t]
 \begin{tikzpicture}[scale=1.1]
\draw[step=1.cm,color=gray!50] (0,0) grid (4,5);
\begin{scope}[shift={(-1,-1)}]
\draw[ line width=.4mm,->] ( 2,2)-- (2.5,2);
\draw[ line width=.4mm] ( 2.4,2)-- (3,2);
\node[scale=.7,below] at ( 2.52,1.9){$U(\vec{n}\,;\hat{1})$};
\node[scale=.7,above] at ( 2,2.07){$E^{\rm a}_+(\vec{n}\,;\hat{1})$};
\node[scale=.7,above] at ( 3.27,2.07){$E^{\rm a}_-(\vec{n}+\hat 1;\hat{1})$};
\node[scale=.7,below] at ( 1.87,2){$\vec{n}$};
\node[scale=.7,below] at ( 3.27,2){$\vec{n}+\hat 1$};
\node[scale=.6,rotate=45] at (2.2,2) {$\blacksquare$};
\node[scale=.6,rotate=45] at (2.8,2) {$\blacksquare$};
\end{scope}
\begin{scope}[shift={(0,2)}]
\draw[ line width=.4mm,] (2,1) --(3,1.);
\draw[ line width=.4mm,] (2,1) --(1,1.);
\draw[ line width=.4mm,] (2,1) --(2,2);
\draw[ line width=.4mm,] (2,1) --(2,.0);
\node[scale=.6,rotate=45] at (2.4,1) {$\blacksquare$};
\node[scale=.6,rotate=45] at (1.6,1) {$\blacksquare$};
\node[scale=.6,rotate=45] at (2,1.4) {$\blacksquare$};
\node[scale=.6,rotate=45] at (2,.6) {$\blacksquare$};
\node[above,scale=.65] at (1.5,1.05) { $E^{\rm a}_-(\vec n; \hat{1})$};
\node[above,scale=.65] at (2.6,1.05) { $E^{\rm a}_+(\vec n; \hat{1})$};
\node[right,scale=.65] at (2,1.65) { $E^{\rm a}_+(\vec n; \hat{2})$};
\node[right,scale=.65] at (2.05,.5) { $E^{\rm a}_-(\vec n; \hat{2})$};
\node[scale=.7] at (1.65,.75) { ${\cal G}^{\rm a}(\vec n)$};
\node[] at (2.,-2.5) { (a)};
\end{scope}
\begin{scope}[shift={(5,0)}]
\draw[step=1.cm,color=gray!50] (0,0) grid (4,5);
\node[below, scale=.7] at (2.5,2.5){ $\mathcal{W}(m,n)$};
\draw[ line width=.4mm,->] (0,0)--(1.14,0.0);
\draw[ line width=.4mm,->] (1,0)--(2.14,0.)--(2.14,1.4);
\draw[ line width=.4mm,->] (2.14,1.4)--(2.14,2)--(3,2)--(3,2.5);
\draw[ line width=.4mm,->] (3,2.5)--(3,3)--(2.5,3);
\draw[ line width=.4mm,->] (2.5,3)--(2,3)--(2,2.4);
\draw[ line width=.4mm,->] (2,2.5)--(2,.8);
\draw[ line width=.4mm] (2,1)--(2,.14)--(0,.14);
\node[scale=.6] at (3.25,3.15){$(m,n)$};
\draw[fill](2.5,2.5) circle (1pt);
\draw[fill](0,.15) circle (1.6pt);
\draw[fill](0,.) circle (1.6pt);
\node[above,scale=.65] at (0.4,.15){$\mathcal{E}^{\rm a}_+(m,n)$};
\node[below,scale=.65] at (0.4,-.02){$\mathcal{E}^{\rm a}_-(m,n)$};
\node[] at (1.9,-.5) {(b)};
\end{scope}
\begin{scope}[shift={(10,0)}]
\draw[step=1.cm,color=gray!50] (0,0) grid (4,5);
\draw[line width=.4mm,->] ( 0, 0)--(1.5,0);
\draw[line width=.4mm] ( 0,0)--(2,0); %
\draw[line width=.4mm] (2,0)--(2,3);
\draw[line width=.4mm,->] (2,0)--(2,1.5);
\draw[fill, gray] (0,0) circle (1.5 pt);
\draw[fill,gray] (2,3) circle (1.5 pt);
\node[below, scale=.7] at (2.48,2.4){ $\mathsf{T}(m,n)$};
\node[above, scale=.7] at (2.45,3.){ $\mathsf{E}^{\rm a}_+(m,n)$};
\node[above,scale=.7] at (0.41,.02){$\mathsf{E}^{\rm a}_-(m,n)$};
\node[] at (1.5,-.5) {(c)};
\end{scope}
\end{tikzpicture}
\caption{(a) Kogut Susskind link formulations. Link operator $U(\vec{n};\hat{i})$ and its left (right) $E^{\rm a}_+(\vec{n};\hat{i})$ ($E^{\rm a}_-(\vec{n}+\hat i;\hat{i})$) electric field. Gauss operator site at $\vec n$, $\mathcal{G}^{\rm a}(\vec{n})= \sum_{i=1}^2 [E^{\rm a}_+(\vec{n};\hat{i})+E^{\rm a}_-(\vec{n};\hat{i})]$ is also shown, (b) Dual physical plaquette holonomy $\mathcal{W}(\vec{n})$ and  their  left (right) $\mathcal{E}_+(\vec{n})$ ($\mathcal{E}_-(\vec{n})$) electric field, (c) Unphysical string holonomy $\mathsf{T}(m,n)$ and it's left $\mathsf{E}_+(\vec{n})(=\mathcal{G}(\vec{n}))$ and right ($\mathsf{E}_-(\vec{n})$) electric field respectively. These string holonomies decouple on physical Hilbert space.}\label{DT}
\end{figure*}
\noindent Also, the relation (\ref{lrrelxx}) implies that their magnitudes are equal: 
\bea
 {\vec{E}_+}^{\,2}(n,\hat i) =  {\vec{E}_-}^{\,2}(n,\hat i; \hat i) \equiv {\vec{ E}}^{\,2}(n,\hat i).
\label{ident22d} 
\eea
It is convenient to represent the independent conjugate operators on a link $l$ by $( E_+(l),\;U_{\alpha\beta}(l))$ or $(E_-(l),\;U_{\alpha\beta}(l))$ as  shown in Figure \ref{DT}-a. They are the initial (before duality) electric variables representing the SU(N) electric fields $E(l)$ and their canonical conjugate  magnetic vector potentials $U(l)$ on the link $l$. 
The SU(N) gauge transformations are 
\begin{align}
\begin{aligned}\label{EU_trans}
U(\vec n;\hat i)\rightarrow &~\Lambda(\vec n)\,U(\vec n;\,\hat i\,)\Lambda^\dagger(\vec n+\hat{i})\\
E_\pm (\vec n;\,\hat i\,)\rightarrow&~ \Lambda(\vec n)\,E_\pm(\vec n;\,\hat i\,)\Lambda^\dagger(\vec n)
\end{aligned}
\end{align}
The generators of  gauge transformation at site $\vec n$ are Gauss operators defined by  
\begin{equation}\label{glc}
\mathcal{G}^{\rm a}(\vec{n})= \sum_{i=1}^{2}\left(E_+^{\mathrm a}(\vec n;\hat i)+E_-^{\mathrm a}(\vec n;\hat i)\right)
\end{equation}
In our earlier work \cite{mansre}, using canonical transformations in $(2+1)$ dimension,  
we solved the  Gauss law constraints
(\ref{glc}) 
\bea 
\mathcal{G}^{\rm a}(\vec{n})= 0, ~~~~\forall ~~\vec n \neq (0,0),  
\label{glc1} 
\eea 
to write down the SU(N) Kogut-Susskind Hamiltonian as a dual SU(N) spin model. 
We summarize the essential results required for the present work in the next section. 

\section{Duality \& Loops}
\label{dts}

In our previous work \cite{mansre}, we obtained exact duality transformations through a series of canonical transformations over the entire lattice in $(2+1)$ dimension. 
The  dual model is 
written  in terms of the  
mutually independent plaquette loops (see Figure \ref{DT}-b) or scalar magnetic flux operators
${\cal W}(p)$ and their conjugate electric scalar potential $\vec {\cal E}(p)$ operators satisfying (\ref{dccr}). The advantage of iterative canonical transformation is that the canonical commutation relations are preserved at every stage \cite{mansre} leading to the exact canonical magnetic description at the end. Note that the dual operators are defined on the  plaquettes or dual lattice sites while the initial Kogut-Susskind operators, discussed in the previous section, are defined on the lattice links. Such dual magnetic description has been useful in the past to study compact U(1)  
and SU(N) lattice gauge theories in $(2+1)$ and $(3+1)$ dimension \cite{mmar,mansre,helleretc}. 
The  dual SU(N) physical and unphysical operators   \cite{mansre}
are summarised in the following two subsections respectively.    

\vspace{0.1cm}
\begin{center}
{\bf 1. Magnetic flux operators}
$(\mathcal{W}_{\alpha\beta}(p),\;\mathcal{E}^{\rm a}_{+}(p))$ 
\end{center}
\vspace{0.1cm}

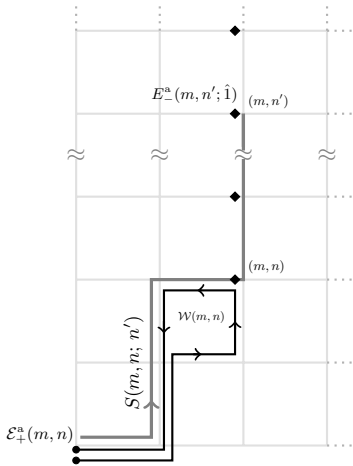
\begin{figure}
    \centering
    \hspace{-.65cm}
\begin{tikzpicture}[scale=1.1]
\begin{scope}[shift={(8,0)}]

\draw[thick,gray!60,opacity=.4] (0,0) grid (3,5);
\foreach [count=\i, evaluate=\i as \x using int(\i+1)] \y in {0,1,2,3,4,5}{
\draw [gray!60,dotted,thick] (3,\y)--(3.35,\y);
}

\foreach [count=\i, evaluate=\i as \x using int(\i+1)] \y in {0,1,2,3}{
\draw [gray!60,dotted,thick] (\y,5)--(\y,5.35);
}
\draw[thick](0,-.18)--(1.15,-.18)--(1.15,1.1)--(1.9,1.1)--(1.9,1.87)--(1.05,1.87)--(1.05,-.05)--(0,-.05);
\draw [fill](0,-.18) circle (.04cm);
\draw [fill](0,-.05) circle (.04cm);
\draw [thick,->](1.9,1.4)--(1.9,1.5);
\draw [thick,->](1.4,1.1)--(1.5,1.1);
\draw [thick,->](1.5,1.87)--(1.45,1.87);
\draw [thick,->](1.05,1.5)--(1.05,1.35);
\node [left,scale=.65] at (2,4.24){$E_-^{\rm a}(m,n';\hat{1})$};
\draw[very thick,color=gray](0.05,.1)--(.9,.1)--(.9,2)--(2,2)--(2,4);
\draw[very thick,color=gray,->](.9,.5)--(.9,.55);
\draw[very thick,color=gray](2,3.6)--(2,4);
\draw [white,fill] (-.1,3.46) rectangle (3.1,3.55);
\node[gray] at (0,3.5) { $\approx$};
\node[gray] at (1,3.5) { $\approx$};
\node[gray] at (2,3.5) { $\approx$};
\node[gray] at (3,3.5) { $\approx$};
\node [right,scale=.55] at (2,4.15){($m,n'$)};
\node [right,scale=.55] at (2,2.15){($m,n$)};
\node [rotate=90,above,scale=.8] at (.89,1){$S(m,n;\,n')$};
\node [scale=.5]at (1.5,1.55){${\cal W}(m,n)$};
\node [below,scale=.7]at (-.43,.3){${\cal E}^{\rm a}_+(m,n)$};
\foreach \y in {2,3,4,5}{
\node[scale=.4,rectangle,draw,rotate=45,fill] at (1.9,\y) {};
}
\end{scope}
\end{tikzpicture}
\caption{Graphical representation of canonical relation (\ref{edt}). We have used ${\scriptstyle \blacklozenge }$ to represent Kogut-Susskind electric fields $E^{\rm a}_-(m,n';\hat{1})$ and $\bullet$ to represent new plaquette electric fields ${\cal E}^{\rm a}
_+(m,n)$. The thick grey line represents parallel transport $  S(m,n;\,n')$ defined in equation (\ref{sungl}).}
    \label{fig:dual_relation}
\end{figure}

They are the physical magnetic operators which solve the SU(N) Gauss law constraints and define the physical Hilbert space ${\cal H}^{\text {phys}}$.
They represent the scalar SU(N)  magnetic fluxes $({\cal W}(p))$ on plaquette $p$ and their conjugate electric scalar potentials ${\cal E}_\pm(p)$ \footnote{The convention chosen for loop (string) electric fields is that $ \mathcal{E}^{\rm a}_{-}(\vec n) ~(\mathbb{E}^{\rm a}_{-}( \vec n))$ and $\mathcal{E}^{\rm a}_{+}(\vec n)~(\mathbb {E}^{\rm a}_{+}(\vec n))$ are  located  at  the  initial,  end  points of  the  flux loop (string) respectively.}.    
The SU(N) duality relations are 
\bea
\hspace{-0.18cm} 
~~~~{\cal W}(m,n) =  \mathsf{T}(m-1,n-1) ~U_p(m,n)~ \mathsf{T}^\dagger(m-1,n-1) ~~\nonumber \\
{\cal E}_+(m,n) \! =\!\sum_{n'=n}^{\infty} \!S^\dagger(m,n;n') E_-(m,n';\hat 1) S (m,n;n')
\phantom{xxxx} 
\label{edt}
 \eea
 The parallel transport operators  $\mathsf{T}(m-1,n-1)$ and $S (m,n;n')$ are defined as (see Figure \ref{DT}-c and Figure \ref{fig:dual_relation}) are 
 \begin{subequations}
             \begin{align} 
       \mathsf{T}(m,n)  =& \;\prod_{m'=0}^m U(m',0\,;\,\hat{1})~\prod_{n'=0}^n \,U(m,n';\,\hat{2}),\label{sunstring}\\
\hspace{-0.18cm}S(m,n;n^\prime)  \equiv & \;{\sf T}(m\hspace{-0.1cm}-\hspace{-0.1cm}1,n)\;U(m\hspace{-0.1cm}-\hspace{-0.1cm}1,n;
   \hat{1})\;\prod_{h=n}^{n'}~U(m,h; \hat{2}).
                 \label{sungl}                 
                \end{align}
                \end{subequations}
Like in the Kogut-Susskind approach, the right electric  potentials  are defined by 
\bea 
{\cal E}_-(p) = -{\cal W}^\dagger (p)\, {\cal E}_+(p)\, {\cal W}(p),\label{lrreld} 
\eea
Note that ${\cal E}^{\rm a}_-(p)$ are attached to the initial end of plaquette flux line ${\cal W}(p)$ as shown in Figure \ref{DT}-b. 
 The dual operator  commutation relations are \cite{mansre}
\begin{align} 
\begin{aligned} 
\left[{\cal E}_+^{\rm a}(p), {\cal W}_{\alpha\beta}(p)\right] =&\; ~~~\left(T^{\rm a} {\cal W}(p)\right)_{\alpha\beta}, \\
\left[{\cal E}_-^{\rm a}(p), {\cal W}_{\alpha\beta}(p)\right] =&\; - \left({\cal W}(p)\;T^{\rm a}\right)_{\alpha\beta}.\label{dccr}
\end{aligned}
\end{align}
The above commutation relations imply that ${\cal E}_+^{\rm a}(p)$ (${\cal E}_-^{\rm a}(p)$) rotate ${\cal W}_{\alpha\beta}(p) $ from left ( right) and therefore are the left (right) electric scalar potentials. They are
mutually independent and satisfy SU(N) algebra:  
\begin{align}
\begin{aligned}
\left[{\cal E}^{\rm a}_+(p), \; {\cal E}^{\rm b}_-(p)\right] =& \;0,\\
\left[{\cal E}^{\rm a}_\pm(p), \; {\cal E}^{\rm b}_\pm(p)\right]=& \;i\,f^{{\rm a}{\rm b}{\rm c}} {\cal E}^{\rm c}_\pm(p).
\label{ecr}
\end{aligned}
\end{align}
Also, the relation (\ref{lrreld}) implies that their magnitudes are equal: 
\bea
 {\vec{\cal E}_+}^{\,2}(p) =  {\vec{\cal E}_-}^{\,2}(p) \equiv {\vec{\cal E}}^{\,2}(p),
\label{ident1d} 
\eea
In the first two equations above we have defined ${\vec{\cal E}_\pm}^{\,2}(p) \equiv 
\sum_{{\rm a}=1}^{\text{N}^2-1} {\cal E}_\pm^{\rm a}(p){\cal E}_\pm^{\rm a} (p)$. The relations  (\ref{lrreld}), (\ref{dccr}), (\ref{ecr}) and (\ref{ident1d}) in this (dual) magnetic  formulation are exactly analogous to the initial 
relations (\ref{lrrelxx}), (\ref{com_EU}), (\ref{com_EE}) and (\ref{ident22d})  respectively in the 
original Kogut-Susskind electric formulation. 
The dual spin or magnetic flux operators  transform as SU(N) adjoint matter field at the origin
\begin{align}
\begin{aligned}\label{gtatorigin}
\mathcal{W}(m,n)\rightarrow &~\Lambda(0,0)\,\mathcal{W}(m,n)\Lambda^\dagger(0,0)\\
\mathcal{E}_\pm (m,n)\rightarrow&~ \Lambda(0,0)\,\mathcal{E}_\pm(m,n)\Lambda^\dagger(0,0)
\end{aligned}
\end{align} 
The canonical transformations (\ref{edt}) can also be easily inverted  to give the Kogut-Susskind electic fields in terms of the dual electric scalar potentials  \cite{mansre}. These inverse relations will not be discussed as they are not relevant for the present work. 

\vspace{0.3cm}
\begin{center}
{\bf 2. String operators} $(\mathsf{E}^{\rm a}_{-}(\vec n),\;\mathsf{T}(\vec n))$
\end{center}
\vspace{0.3cm}

They are unphysical operators and represent SU(N) gauge degrees of freedom 
at every lattice site away from the origin. 
They are shown in Figure \ref{DT}-c.  
\begin{align} 
\begin{aligned} 
\mathsf{T}(m,n)  =\;\prod_{m'=0}^mU(m',0;\,\hat{1})~\prod_{n'=0}^n \,U(m,n';\,\hat{2}),\\
{\mathsf E}^{\rm a}_+(m,n)  =\;\mathcal{G}^{\rm a}(m,n) \simeq 0.~~~~~~~~ ~~~~~
\end{aligned}\label{sdt} 
\end{align} 

Thus all string operators ${\mathsf T}(m,n)$ become cyclic as their conjugate electric fields ${\mathsf E}^a_+(m,n)$  turns out to be the Gauss law operator ${\cal G}^a(m,n)$ \cite{mansre}. Therefore they vanish on the physical Hilbert space ${\cal H}^p$ where the SU(N) Gauss laws are satisfied. The string operators, being unphysical,  will not be relevant in this work and will not be considered henceforth.
\section{Disorder operators}
\label{sundo} 

As mentioned earlier, the order and disorder operators in SU(N) lattice theory are 
simply the shift or the creation-annihilation  operators for the gauge invariant electric and magnetic fluxes respectively. Note that the
Wilson loop operators ${\cal W}^{[\vec j]}({\cal C})$, constructed  in terms of the magnetic vector potentials $U(l)$ in (\ref{wlo}), shift their conjugate electric fluxes along the loop ${\cal C}$. 
In this section, we construct the gauge invariant disorder operators which are dual to the Wilson loop operators ${\cal W}^{[\vec j]}({\cal C})$ and shift the magnetic fluxes instead. For the sake of simplicity, we first consider SU(2) case and then generalize it to SU(3) and finally to the SU(N) gauge group. All the algebraic details for the SU(N) electric \& magnetic basis are given in Appendices  \ref{elb} and \ref{mlb} respectively. 

\subsubsection[1]{\bf{SU(2) Disorder Operator}}
\label{su2disop}

 \noindent The magnetic plaquette flux operator, 
 \bea 
 {\cal W}^{[j=\frac{1}{2}]}(p) \equiv \exp\frac{i}{2} \left(\hat n(p) \cdot \vec \sigma \;\; \omega(p)\right). 
 \label{su2m}
 \eea 
 can also be rewritten in the  angle-axis representation as: 
 \begin{align}
 \begin{aligned} 
{\cal W}^{[j=\frac{1}{2}]}{(p)} &\equiv \cos\left(\frac{\omega{(p)}}{2}\right) \sigma_{0} + i \left(\hat{n}{(p)} \cdot \vec \sigma\right) \sin\left(\frac{\omega{(p)}}{2}\right); \\
 	 &~~~\hat n{(p)}\cdot \hat n{(p)} =1,~~~ \forall ~(p).  \label{wmn}
 	\end{aligned} 
  \end{align}
 In (\ref{wmn}) $\sigma_0, \vec \sigma (\equiv \sigma_1,\sigma_2,\sigma_3)$ are  the unit, 3 Pauli matrices respectively. Under global gauge transformation $\Lambda \equiv \Lambda(0,0)$ in (\ref{gtatorigin}), $(\omega,\hat n)$  transform as: 
 \begin{align} 
 \begin{aligned} 
  \omega(p)&\rightarrow \omega(p),\\ 
  \hat n(p) \equiv \sum_{{\rm a}=1}^{3} \hat n^{\rm a}(p)\sigma^{\rm a} & \rightarrow \Lambda(0,0) ~\hat n(p)~ \Lambda^\dagger(0,0). 
  	\label{gtp}
   \end{aligned} 
 \end{align}
 Thus  $\omega(p)$ are gauge invariant angle and  $\hat n(p)$ are the unit vector operators.  We now define  two unitary operators: 
 \begin{align}
\begin{aligned}
 {\Sigma }^{+}_\theta(p) &\equiv  \exp ~i \Big( \;{\hat n }(p)\cdot {\cal E}_{+}(p)\; {\theta}\phantom{.}  \Big)	\\
 {\Sigma }^{-}_\theta(p) &\equiv \exp ~i \Big(\;{\hat n }(p)\cdot {\cal E}_{-}(p) \;{\theta}\phantom{.}  \Big),
 \label{mfo}
 \end{aligned}
 \end{align} 
 which are located on a plaquette $p$.  They both are gauge invariant because ${\cal E}_{\pm}^{\rm a}(p)$ and  $\hat n (p)$ gauge  transform like vectors as shown in (\ref{gtatorigin}) and (\ref{gtp}). In other words, $\left[{\cal G}^{\rm a}, \Sigma^{\pm}_\theta(p)\right] =0$, where ${\cal G}^{\rm a}$ is defined  in (\ref{glc}). As the left and right electric scalar potentials  are related through 
 (\ref{lrreld}),  ${\Sigma}^{\pm}_\theta(p)$ are not mutually independent and satisfy  
 \footnote{ We have used the relation $n^{\rm a}(p)\mathcal{E}^{\rm a}_-(p) =- n^{\rm a}(p) R^{\rm ab}(\mathcal{W}^\dagger(p))\mathcal{E}^{\rm b}_+(p)$ and 
 \vspace{-0.11cm}
  \begin{align*}
 \phantom{XX} n^{\rm a}(p) R^{\rm ab}&(\mathcal{W}^\dagger(p))\\&=\text{Tr} (\sigma^{\rm a} \mathcal{W}(p))\;\frac{1}{2}\text{Tr}(\sigma^{\rm a} \mathcal{W}^\dagger(p) \sigma^{\rm b}\mathcal{W}(p))\\
  &=\;\frac{1}{2}(\sigma^{\rm a}_{\alpha\beta}\sigma^{\rm a}_{\gamma\delta}) \sigma^{\rm b}_{\eta \rho}      \mathcal{W}_{\beta\alpha}(p)\mathcal{W}^\dagger_{\delta\eta}(p)\mathcal{W}_{\rho\gamma}(p)\\
  &=\;\frac{1}{2}(2\delta_{\alpha\delta}\delta_{\beta\gamma} -\delta_{\alpha\beta}\delta_{\gamma\delta})\sigma^{\rm b}_{\eta \rho}\mathcal{W}_{\beta\alpha}(p)\mathcal{W}^\dagger_{\delta\eta}(p)\mathcal{W}_{\rho\gamma}(p)\\
  &= \sigma^{\rm b}_{\eta \rho}\mathcal{W}_{\rho\eta}(p)=n^{\rm b}(p)\label{nrn}
 \end{align*}}: 
 \begin{align} 
 	{\Sigma}^{+}_{\theta}(p) ~{\Sigma}^{-}_{\theta}(p) =
 	{\Sigma}^{-}_{\theta}(p) ~{\Sigma}^{+}_{\theta}(p)
 	= {\cal I}. 
  \label{ssdi}
 \end{align}
 In (\ref{ssdi}),  ${\cal I}$ denotes the unit operator in the physical Hilbert space ${\cal H}^p$.  The identities (\ref{ssdi}) can be easily obtained by using ${\cal E}_-(p) = -R^{\rm  ab}(\hat n,\omega) {\cal E}_+(p)$  and $R^{\rm ab}(\hat{n},\omega)\hat{n}^{\rm b} = \hat{n}^{\rm a}$ where  $R^{{\rm a}{\rm b}}(\hat n, \omega) =  \frac{1}{2} \text{Tr}(\sigma^{\rm a}{\cal W} \sigma^{\rm b} {\cal W}^\dagger)$. 

\vspace{0.5cm} 
\begin{center} 
{\it{A. SU(2) Prepotential Operators}}
\label{su2ppo} 
\end{center} 
\vspace{0.5cm} 

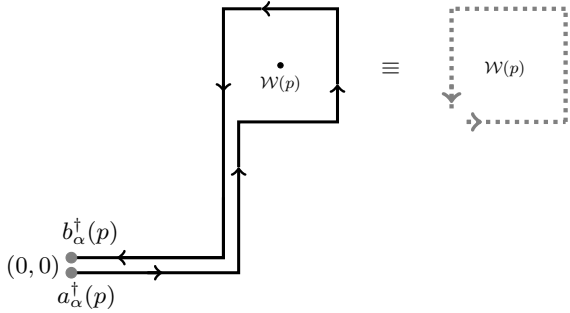
\begin{figure}
    \centering
    \begin{tikzpicture}
\begin{scope}[shift={(5,0)}]
\node[ scale=.7] at (2.75,2.5){ $\mathcal{W}(p)$};
\draw[ line width=.4mm,->] (0,0)--(1.2,0.0);
\draw[ line width=.4mm,->] (1,0)--(2.2,0.)--(2.2,1.4);
\draw[ line width=.4mm,->] (2.2,1.4)--(2.2,2)--(3.5,2)--(3.5,2.5);
\draw[ line width=.4mm,->] (3.5,2.5)--(3.5,3.5)--(2.5,3.5);
\draw[ line width=.4mm,->] (2.5,3.5)--(2,3.5)--(2,2.4);
\draw[ line width=.4mm,->] (2,2.5)--(2,.2)--(.6,.2);
\draw[ line width=.4mm] (2,1)--(2,.2)--(0,.2);
\draw [gray,dotted, ultra thick] (5.2,2)--(6.5,2);
\draw [gray,dotted, ultra thick, ->] (5.2,2)--(5.4,2);
\draw [gray,dotted, ultra thick] (6.5,2)--(6.5,3.5);
\draw [gray,dotted, ultra thick] (6.5,3.5)--(5,3.5);
\draw [gray,dotted, ultra thick, -] (5,3.5)-- (5.,2.15);
\draw [gray,dotted, ultra thick, ->] (5,2.5)-- (5.,2.25);
\node[below,scale=1.] at (4.2,2.9){$\equiv$};
\node[ scale=.7] at (5.7,2.7){ $\mathcal{W}(p)$};
\draw[fill](2.75,2.75) circle (1pt);
\draw[fill,gray](0,.2) circle (2pt);
\draw[fill,gray](0,.) circle (2pt);
\node[above,scale=1] at (0.25,.25){$b^{\dagger}_{\alpha}(p)$};
\node[below,scale=1.] at (0.2,.0){$a^{\dagger}_\alpha(p)$};
\node[below,scale=1.] at (-0.5,.35){$(0,0)$};
\end{scope}
    \end{tikzpicture}
    \caption{SU(2) prepotential operators in the dual formulation: The two ends of the plaquette flux operator ${\cal W}(p)$ are associated with two doublets of the harmonic oscillators at the origin $(0,0)$ \cite{mansre,manpp}. Under gauge transformations at the origin, $\left(a^\dagger_\alpha(p), b^\dagger_\beta(p)\right)$  transform as SU(2) doublets. The dotted plaqette on the right hand side is a compact way to represents the plaquette holonomy ${\cal W}(p)$.}
    \label{fig:boson}
\end{figure}

\noindent It is extremely convenient to use the prepotential \cite{mansre,manpp} representation for the dual electric  potential on the plaquette loops to construct the electric loop (Appendix \ref{elb}) as well the magnetic loop (Appendix \ref{mlb}) basis. This simplification is illustrated in Figure \ref{fig:boson}. A further advantage is that this simple procedure can be directly generalized to  all SU(N). We write the SU(2) dual plaquette loop electric potentials on  any plaquette $p$ satisfying (\ref{ecr})  as \footnote{In defining ${\cal E}_-^a(p)$ we have used the fact that like (${\sigma}^a$), their transpose with a negative sign ($-\tilde{\sigma}^a$) also satisfies the same SU(2) Lie algebra.}
\begin{align}
\begin{aligned} 
{\cal E}^{\rm a}_+(p) ~&\equiv& a^\dagger(p) \frac{\sigma^{\rm a}}{2} a(p); \\  {\cal E}^{\rm a}_-(p) ~&\equiv& -b(p) \frac{\sigma^{\rm a}}{2} b^\dagger(p).
\label{dppp}
\end{aligned} 
\end{align} 
In (\ref{dppp}), $a^\dagger_\alpha(p)$ and $ b^\dagger_\alpha(p)$ are the two mutually commuting SU(2) doublets of harmonic oscillator creation operators  on every plaquette loop. The standard commutation relations are 
\begin{align} 
\begin{aligned} 
\left[a_\alpha(p),a^\dagger_\beta(p')\right] & = &\delta_{pp'} \delta_{\alpha\beta}, \\
\left[b_\alpha(p),b^\dagger_\beta(p')\right] & = &\delta_{pp'} \delta_{\alpha\beta}. 
\label{ppccr} 
\end{aligned} 
\end{align} 
Using (\ref{ppccr}), it is easy to check that the representation (\ref{dppp}) satisfies (\ref{ecr}). 
The constraints (\ref{ident1d}) imply that 
\bea 
N(p) \equiv a^\dagger(p)\cdot a(p)= b^\dagger(p)\cdot b(p). 
\eea 
The plaquette holonomy in this representation is \cite{manpp}
\bea 
\hspace{-0.2cm} {\cal W}_{\alpha\beta}(p) = F(N) \left[ b^\dagger_\alpha(p)  a^\dagger_\beta(p) + \tilde{b}_\alpha \tilde{a}_\beta \right] F(N) 
\label{plaqhol} 
\eea
 In (\ref{plaqhol}) $F(N) \equiv \frac{1}{\sqrt{(N(p)+1)}}$ is the normalization factor and $\tilde x_\alpha \equiv \epsilon_{\alpha\beta} x_\beta$. The harmonic oscillator representation (\ref{dppp}) implies that $a^\dagger_\alpha$ and  $b^\dagger_\alpha$ transform like doublet from right and anti-doublet from left respectively 
 on every plaquette (p):
\begin{align} 
\begin{aligned} 
\left[{\cal E}_+^{\rm a}(p), a^\dagger_\alpha(p)\right] ~&= \;- \left(a^\dagger(p) \,\frac{\sigma^{\rm a}}{2}\right)_\alpha, \\ 
\left[{\cal E}_-^{\rm a}(p), b^\dagger_\alpha(p)\right]~ &=  \;~~\left(\frac{\sigma^{\rm a}}{2}\, b^\dagger(p)\right)_\alpha. 
\label{su2ptr}
\end{aligned} 
\end{align}
The strong coupling vacuum on every plaquette in the dual formulation  $\ket{0}_p$  satisfies:  \bea 
{\cal E}^{\rm a}_\pm(p) \;\ket{0}_p =0, ~~~\forall p.
\label{dscv2} 
\eea 
This is equivalent to demanding 
\bea 
a_\alpha(p) \ket{0}_p =0, ~~~ b_\alpha(p) \ket{0}_p =0. \label{ab00}
\eea 
The relations (\ref{su2ptr}) and (\ref{ab00}) will be useful to study the action of SU(2) disorder operators on the magnetic basis discussed below. Note that under SU(2) gauge transformations (\ref{gtatorigin}) with $\Lambda(0,0)$ at the origin (see Figure \ref{DT}-b) these oscillators 
transform doublets:
\begin{align}
\begin{aligned}
    a^\dagger_\alpha(p) \rightarrow & ~~   a^\dagger_\beta(p)~ \Lambda_{\beta\alpha}(0,0), ~~~~~~\forall p,  
\\
 b^\dagger_\alpha(p) \rightarrow & ~~   \Lambda^\dagger_{\alpha\beta}(0,0) \; b^\dagger_\beta(p) ~~~~~~~~\forall p. 
 \label{su2gto} 
 \end{aligned} 
\end{align}
 These relations are useful to construct the gauge invariant operators in the magnetic basis constructed in the next section.  
 \vspace{0.5cm} 
\begin{center}
 {\it{B. ~SU(2) Magnetic Basis}}
 \end{center}
 \vspace{0.3cm} 
 \noindent The physical meaning of the operators $\Sigma_\theta^{\pm}(p)$  is simple. The non-Abelian electric scalar potentials  ${\cal E}^{\rm a}_{\pm}(p)$ are conjugate to the magnetic flux operators ${\cal W}^{[j=\frac{1}{2}]}_{\alpha\beta}(p)$. They satisfy the canonical commutation relations (\ref{dccr}). Therefore, the gauge invariant vortex operator operator ${\Sigma}^{\pm}_{\theta}(p)$ acting on the magnetic basis on a plaquette  changes  the magnetic flux on it continuously as a function of $\theta$ in (\ref{su2oda}). To see this explicitly, we first construct the SU(2) magnetic basis.  
 We note that 
 $$\left[{\cal W}_{\alpha\beta}(p),{\cal W}_{\gamma\delta}(p')\right] =0, ~~\forall\; p,p'.$$ 
 Therefore we can diagonalize all 4 operators $\left({\cal W}_{11}(p), {\cal W}_{12}(p), 
{\cal W}_{21}(p), {\cal W}_{22}(p)\right)$ simultaneously on every plaquette. The common eigenstates 
$\ket{Z(p)}\equiv \ket{z_1(p),z_2(p)}$  satisfy 
\bea 
{\cal W}_{\alpha\beta}(p)~ \ket{Z(p)} = Z_{\alpha\beta}(p) ~\ket{Z(p)}, ~~~~\alpha,\beta =1,2. 
\label{su2eve}
\eea
 In (\ref{su2eve}) the SU(2) matrix on the plaquette $p$ is 
 \bea 
 Z =\left(   \begin{array}{c}
 \!\!\!\!~~~z_1\, ~~~~~~~z_2\!\!\\
  -z_2^*\; ~~~~~~z_1^*
    \end{array} \right), ~~~~|z_1|^2+|z_2|^2 =1.
    \label{su2zm}
\eea 
The SU(2) $Z$ matrices can also be written in  the SU(2) angle-axis representation 
 \bea
Z = e^{i\, \omega\, \hat n^{\rm a} \,\frac{\sigma^{\rm a}}{2}}.   
 \label{su2fo}
 \eea
 The two SU(2) representations (\ref{su2zm}) and (\ref{su2fo}) are related by 
 $$z_1=\cos\left(\frac{\omega}{2}\right)+i \hat n^3\sin\left(\frac{\omega}{2}\right), ~ z_2 =(\hat n^2+i \hat n^1) \sin\left(\frac{\omega}{2}\right).$$
We now construct $\ket{Z(p)}$ and show that on this basis the vortex operator $\Sigma^\pm_{\theta} (p)$
act as  the shift operators for the plaquette magnetic fluxes. The magnetic eigenstates  $\ket{Z(p)}$ can be explicitly constructed in terms of SU(2) prepotential operators \cite{manpp} (see Appendix \ref{mlb}):
 \begin{align}
 \ket{Z(p)}= \sum_{j(p)=0}^\infty \sqrt{d(j(p))}\;\frac{\left(a^\dagger(p)\; Z(p)\; b^\dagger(p)\right)^{2j(p)}}{(2j(p))!}\;\ket{0}_p.
 \label{xvc}
\end{align}
In (\ref{xvc}) 
$d(j) \equiv (2j+1)$ is the dimension of the $j$ representation and 
$(a^\dagger \,Z\, b^\dagger) \equiv \sum_{\alpha,\beta=1}^2 \left( a^\dagger_\alpha \, Z_{\alpha\beta} \; b^\dagger_\beta\right )$.  
From now onwards we will ignore the  plaquette index $p$ on all the operators and the states as they are all defined on the lattice plaquettes. The magnetic eigenstates (\ref{xvc}) have simple SU(2) 
gauge transformation properties
\bea 
\ket{Z} \rightarrow \ket{\Lambda Z \Lambda^\dagger}, ~~~\Lambda \equiv \Lambda(0,0).  
\label{sgtsu2} 
\eea 
The transformations (\ref{sgtsu2}) are clear from (\ref{su2gto}) and (\ref{xvc}). 
In the angle axis representation (\ref{su2fo}) the  gauge transformations (\ref{sgtsu2}) take simpler form 
 \bea 
 \omega(p) \rightarrow \omega(p), ~~~ \hat n(p) \, \rightarrow \, \Lambda \, \hat n(p) \, \Lambda^\dagger, ~~~~\Lambda \equiv \Lambda(0,0). 
 \eea 
 Thus $\omega(p), ~\forall~ p$ are gauge invariant angles and $\hat n(p)~\forall p$ transform globally like  SU(2) adjoint vectors.  The eigenvalues of the plaquette magnetic field operators in the Hamiltonian (\ref{ham}) are:
 \begin{align} 
 \text{Tr}\left( {\cal W}^{[j=\frac{1}{2}]}\right) ~\ket{Z(\omega,\hat n)}  = 2\cos\left(\frac{\omega}{2}\right)~ \ket{Z(\omega,\hat n)}.	\label{wlev}  
 \end{align}   
Now we evaluate the action of  disorder operator using the prepotential relations
\begin{align} 
\begin{aligned} 
&\Sigma_\theta^{+}a^\dagger_\alpha \Sigma_\theta^{-}=\left(a^\dagger e^{\frac{i}{2} \theta \hat n^{\rm a} \sigma^{\rm a}}\right)_\alpha, \\
&\Sigma_\theta^{-} b^{\dagger}_{\alpha} \Sigma_\theta^{+}=\left( e^{-\frac{i}{2} \theta \hat n^{\rm a} \sigma^{\rm a}} b^\dagger\right)_\alpha.
\label{pptr} 
\end{aligned}
\end{align}
The relations (\ref{pptr}) can be easily established using (\ref{mfo}) and  the 
prepotential representation of $\mathcal{E}_\pm(p)$ in (\ref{dppp}). 
\begin{align}
\begin{aligned}
 \Sigma_\theta^{+} \ket{Z(\omega, \hat n)} \,=& \; \ket{e^{~\frac{i}{2} \theta \hat n^{\rm a} \sigma^{\rm a}}Z(\omega, \hat n)}=\ket{Z(\omega+\theta, \hat n)}\\
  \Sigma_\theta^{-} \ket{Z(\omega, \hat n)} \, =& \;\ket{Z(\omega, \hat n)e^{-\frac{i}{2} \theta \hat n^{\rm a} \sigma^{\rm a}}}=\ket{Z(\omega-\theta, \hat n)}
\end{aligned}\label{aabbcc}
\end{align}
Thus  the SU(2) plaquette disorder operator $\Sigma_\theta^{\pm} $ translates the  plaquette magnetic fluxes. This is precisely dual to the action of the  Wilson loop operators which translate the SU(2) loop electric fluxes as shown in Appendix \ref{elb} (see eqn (\ref{trae}) and Figure \ref{fig:tadpole}).

 \vspace{0.5cm} 
 \begin{center} 
{{\it C. ~SU(2) Order-Disorder Algebra}}
\end{center} 
\vspace{0.3cm} 
The dual canonical commutation relations (\ref{dccr}) involving magnetic plaquette flux operators ${\cal W}(p)$ and their conjugate electric scalar potential ${\cal E}(p)$ immediately lead to the SU(2) order-disorder algebra:
\begin{align}
\begin{aligned}
\Sigma^+_{\theta} (p)\,&\mathcal{W}^{[j=\frac{1}{2}]}_{\alpha\beta}(p) \,\Sigma^{-}_{\theta} (p)= D^{[j=\frac{1}{2}]}_{\alpha\gamma}(\hat n, \theta) \,\mathcal{W}^{[j=\frac{1}{2}]}_{\gamma\beta}(p)\\
\Sigma^-_{\theta} (p)\,&\mathcal{W}^{[j=\frac{1}{2}]}_{\alpha\beta}(p) \,\Sigma^{+}_{\theta} (p)= \mathcal{W}^{[j=\frac{1}{2}]}_{\alpha\gamma}(p)\,D^{[j=\frac{1}{2}]}_{\gamma\beta}(\hat n, \theta).
\label{odal}
\end{aligned}
\end{align}
In (\ref{odal}) the Wigner matrix $D^{[j=\frac{1}{2}]} \equiv  e^{i \; \hat n^{\rm a} \cdot \; \sigma^{\rm a}\; \frac{\theta}{2}}$ are the rotation matrix in $j=\frac{1}{2}$  representation around the magnetic axis $\hat n(p)$ defined through the plaquette loops ${\cal W}(p)$. 
In any higher $[j]$ representation, we can write: 
$${\cal W}^{[j]}_{\alpha\beta} =
  {\cal W}^{[j=1/2]}_{\{\alpha_1\beta_1}{\cal W}^{[j=1/2]}_{\alpha_2\beta_2}\cdots {\cal W}^{[j=1/2]}_{\alpha_{2j}\beta_{2j}\}}$$ 
with all  the $\alpha$ (and therefore $\beta$)  
indices are completely symmetrized. Inserting the disorder operators $(\Sigma)$  and their inverses $(\Sigma^{\dagger})$ in the middle, we get the SU(2) order-disorder algebra relation in $j$ representation.
\begin{align}
\begin{aligned} 
 \Sigma^+_{\theta} (p)\;\mathcal{W}^{[j]}_{\alpha\beta}(p) \;\Sigma^-_{\theta} (p)=&\; D^{[j]}_{\alpha\gamma}(\hat n, \theta) \;
\mathcal{W}^{[j]}_{\gamma\beta}(p),\\
 \Sigma^-_{\theta} (p)\;\mathcal{W}^{[j]}_{\alpha\beta}(p) \;\Sigma^+_{\theta} (p)= & \;
\mathcal{W}^{[j]}_{\alpha\gamma}(p)\,D^{[j]}_{\gamma\beta}(\hat n, \theta).
\label{su2oda}
\end{aligned} 
\end{align}
 In the special case when the rotations are restricted to  the centre $Z_2$ of the SU(2) group then $\theta = 0$ or $2 \pi$ 
in (\ref{su2oda}) and we recover the  't Hooft Wilson order-disorder algebra with $D^{[j]}_{\alpha\beta}(\theta = 2 \pi) = (-1)^{2j}\; \delta_{\alpha\beta}$.
 \begin{align} 
 \Sigma_{\theta=2\pi}^\pm \;  {\cal W}^{[j]}_{\alpha\beta}    = (-1)^{2j} \,  {\cal W}^{[j]}_{\alpha\beta} \;\Sigma_{\theta = 2\pi}^\pm. 	\label{sthr}
 \end{align}
  In (\ref{sthr}), $(-1)^{2j}$ is the n-ality of the j representation. We thus recover the standard Wilson-'t Hooft loop $Z_2$ algebra \cite{hooft, tomb,dis_nonabelian,mack} for SU(2) at $\theta =2\pi$.
  The  operator  $\Sigma_{2\pi}  \equiv  \Sigma^+_{2\pi}    =  \Sigma^-_{2\pi} $ is the  SU(2) 't Hooft operator.

\subsubsection[2]{\bf SU(3) Disorder Operator}
\label{su3disop}

In this section, we construct the disorder operator for SU(3) lattice gauge theory before going to SU(N) gauge group. As in the previous SU(2) case, they are the SU(3) magnetic vortex creation-annihilation operators and are expected to magnetically 
disorder the weak coupling ground state \cite{dis_nonabelian,magnetic_disorder}. 
The SU(3) plaquette magnetic flux operators can be written as  
\bea 
{\cal W}^{[{\textrm p} =1\; {\textrm q}=1]}(p) = 
\exp {i} \left(\hat n(p) \cdot \vec \lambda \;\; \omega(p)\right).
\label{su3fo}
\eea
In (\ref{su3fo}) $\lambda^{\rm a}  ({\rm a}=1,\cdots ,8)$ are the Gell-Mann matrices.
We can also use the angle-axis  representation  \cite{mallesh1997algebra} to write: 
\bea 
{\cal W}^{[{\textrm p} =1\; {\textrm q}=1]}(p) \equiv 
A \, {\cal  I}+ B \,  \vec{n} \cdot \vec \lambda + C \, \vec{n}\star \vec{n} \cdot \vec \lambda.
\label{su3aar}
\eea
In (\ref{su3aar}) $(\vec{n}\star \vec{n})^{\rm a} \equiv d^{{\rm a}{\rm b}{\rm c}}\, \vec{n}^{\rm b} \, \vec{n}^{\rm c}$ 
defines the  second  independent vector with the help of 
 the  SU(3) symmetric tensors $d^{{\rm a}{\rm b}{\rm c}}$.  
 Instead of following the standard polar decomposition (\ref{su3aar}), it is more convenient for us to  construct the two independent SU(3) axes operators as \footnote{The two definitions for 
$\vec{n}_{[1]}(p)$ and $\vec{n}_{[2]}(p)$ in (\ref{Su3axes1}) and (\ref{Su3axes2}) respectively are easily generalizable to SU(N) case discussed in the next section.}
\vspace{-0.25cm}
\begin{subequations}
\begin{align}
\vec{n}^{\rm a}_{[1]}(p)=&\;~\text{Tr}\lambda^{\rm a}\left( \mathcal{W}^{[1,1]}(p)+ \mathcal{W}^{\dagger[1,1]}(p)\right), \label{Su3axes1}\\
\vec{n}^{\rm a}_{[2]}(p)=&~~~~ \;\sqrt{3}\;d^{{\rm a}{\rm b}{\rm c}} \vec{n}^{\rm b}_{[1]}(p)\vec{n}^{\rm c}_{[1]}(p).
\label{Su3axes2}
\end{align}    
\end{subequations}
Note that $\vec{n}^{\rm a}_{[1]}(p), \vec{n}^{\rm a}_{[2]}(p)$ are real. 
Under SU(3) gauge transformations (\ref{gtatorigin}) the above two operators transform as:
\begin{align}
\begin{aligned}
  \vec{n}^{\rm a}_{[1]}(p) \rightarrow &R^{{\rm a}{\rm b}}(\Lambda)\;\vec{n}^{\rm b}_{[1]}(p) \\ 
   \vec{n}^{\rm a}_{[2]}(p) \rightarrow& R^{{\rm a}{\rm b}}(\Lambda)\;\vec{n}^{\rm b}_{[2]}(p).
   \label{arsu3} 
\end{aligned}
\end{align}
In (\ref{arsu3}) $R^{{\rm a}{\rm b}}(\Lambda) =  \frac{1}{2} \text{Tr}(\lambda^{\rm a}\Lambda^\dagger \lambda^{\rm b}\Lambda)$ and $\Lambda \equiv \Lambda(0,0)$. These two axes are linearly independent. It can be  shown that in SU(3) case there exist only two independent axes  as the third axis defined  using another $d^{{\rm a}{\rm b}{\rm c}}$ is the first axis $\vec{n}_{[1]}$ \footnote{ We have used the following two identities\\
     $1.~     (f^{{\rm a}{\rm b}{\rm e}}d^{{\rm d}{\rm c}{\rm e}}+f^{{\rm a}{\rm c}{\rm e}}d^{{\rm b}{\rm d}{\rm e}}+f^{{\rm a}{\rm d}{\rm e}}d^{{\rm b}{\rm c}{\rm e}})=0$\\
    $2.~  (d^{{\rm a}{\rm b}{\rm e}}d^{{\rm d}{\rm c}{\rm e}}+d^{{\rm a}{\rm c}{\rm e}}d^{{\rm b}{\rm d}{\rm e}}+d^{{\rm a}{\rm d}{\rm e}}d^{{\rm b}{\rm c}{\rm e}})=\tfrac{1}{3}(\delta ^{{\rm a}{\rm b}}\delta^{{\rm c}{\rm d}}+\delta ^{{\rm a}{\rm c}}\delta^{{\rm b}{\rm d}}+\delta ^{{\rm b}{\rm c}}\delta^{{\rm a}{\rm d}})$
}:
\begin{align*}
f^{{\rm a}{\rm b}{\rm c}} \vec{n}^{\rm b}_{[2]}(p)\vec{n}^{\rm c}_{[1]}(p)=0,\hspace{2cm}\\   d^{{\rm a bc}} \vec{n}^{\rm b}_{[2]}(p)\vec{n}^{\rm c}_{[1]}(p)=\frac{1}{\sqrt{3}}(\vec{n}^{\rm b}_{[1]}(p)\vec{n}^{\rm b}_{[1]}(p)) ~\vec{n}^{\rm a}_{[1]}(p). 
\end{align*}
Now we define the SU(3) disorder operators which translate these two gauge invariant magnetic fluxes:
\begin{align}
\begin{aligned}
    \Sigma^+_{\theta_1, \theta_2}(p) &\equiv& \exp\left(i  \sum_{h=1}^2  \theta_h (p)\hat{n}^{\rm a}_{[h]}\right)\mathcal{E}^{\rm a}_+(p), \\
    \Sigma^-_{\theta_1, \theta_2}(p) &\equiv& \exp\left(i  \sum_{h=1}^2  \theta_h (p)\hat{n}^{\rm a}_{[h]}\right)\mathcal{E}^{\rm a}_-(p)
    \label{SU(3)do}
\end{aligned}
\end{align}
In (\ref{SU(3)do}) $(\theta_1,\theta_2)\equiv (\theta_1(p),\theta_2(p))$ are the external angular parameters characterizing the SU(3) disorder operator. Like in the SU(2) case, the two operators in (\ref{SU(3)do}) are unitary and  Hermitian conjugate of each other
\begin{equation}
 \Sigma^+_{\theta_1, \theta_2}(p)\; \Sigma^-_{\theta_1, \theta_2}(p) = {\cal I} = 
\Sigma^-_{\theta_1, \theta_2}(p) \; \Sigma^+_{\theta_1, \theta_2}(p). 
\label{ssdi3} 
\end{equation}
Like SU(2) case this can also be proved using the properties of the SU(3) $\lambda$ matrices. 

\vspace{0.2cm}
\begin{center} 
{\it{A. SU(3) Prepotential Operators}}
\end{center}
The SU(3) prepotential operators on plaquettes are defines as 
\begin{align}
\begin{aligned} 
&{\cal E}^{\rm a}_+ \equiv  
~~\sum_{h=1}^{2} \,a^\dagger[h] \,\frac{\lambda^{\rm a}}{2}\, a[h] \\  
&{\cal E}^{\rm a}_-  \equiv  
- \sum_{h=1}^{2} \, b[h] \,\frac{\lambda^{\rm a}}{2}\, b^\dagger[h]. 
\label{dppp3}
\end{aligned} 
\end{align} 
In (\ref{dppp3}), $(a^\dagger_\alpha[h], a_\alpha[h])$ and $(b^\dagger_\alpha[h], b_\alpha[h])$ where $\alpha=1,2,3; h=1,2$ are the  mutually independent SU(3) triplets of harmonic oscillator creation-annihilation  operators  on every plaquette \footnote{We are ignoring the SU(3) multiplicity problem here as the aim in this work is to construct the SU(3) magnetic eigenstates and not to worry about SU(3) multiplicities. One can trivially replace all SU(3) prepotentials by SU(3) irreducible prepotentials \cite{manpp} at the end without changing any results of this section. The same strategy will  be adapted in the next SU(N) section to keep the discussion simple.}.  They are attached to the initial and the end points of the plaquette loops (see Figure \ref{DT}-b). The summation over $[h]=1,2$ is over the rank of the group. As all operators are defined on  plaquettes, we suppress the plaquette index `$p$' throughout this section.  The harmonic oscillator  commutation relations and (\ref{dppp3}) imply that $a^\dagger_\alpha[h]$ and  $b^\dagger_\alpha[h]$ transform like triplets from right and anti-triplets from left respectively on every plaquette ($p$):
transformations: 
\begin{align} 
\begin{aligned} 
\left[{\cal E}_+^{\rm a}, a^\dagger_\alpha[h]\right] ~&= \;~~\left(a^\dagger[h] \,\frac{\lambda^{\rm a}}{2}\right)_\alpha, ~~~~h=1,2, \\ 
\left[{\cal E}_-^{\rm a}, b^\dagger_\alpha[h]\right]~ &=  \;-\left(\frac{\lambda^{\rm a}}{2}\, b^\dagger[h]\right)_\alpha, ~~~~h=1,2. 
\label{su3ptr}
\end{aligned} 
\end{align} 
Like in  SU(2) case (\ref{su2gto}), the SU(3) gauge transformations (\ref{gtatorigin}) with $\Lambda(0,0)$ at the orgin (see Figure \ref{DT}-b) the SU(3) oscillators on every plaquette  
transform as SU(3) triplets:
\begin{align}
\begin{aligned}
    a^\dagger_\alpha[h] \rightarrow & ~~   a^\dagger_\beta[h]~ \Lambda_{\beta\alpha}(0,0), ~~~~h=1,2,  
\\
 b^\dagger_\alpha[h] \rightarrow & ~~   \Lambda^\dagger_{\alpha\beta}(0,0) \; b^\dagger_\beta[h], ~~\,~~h=1,2. 
 \label{su3gto} 
 \end{aligned} 
\end{align}
These relations are again useful for the gauge covariant parametrization of the SU(3) 
magnetic basis in the angle axis representation and is discussed in the next section. 
The SU(3) strong coupling vacuum in the dual description $\ket{0}$ satisfies  
\bea 
{a}_\alpha[h] \;\ket{0}_p =0,~~~{b}_\alpha[h] \;\ket{0}_p =0,~~~ ~~~h=1,2.
\label{dscv2x} 
\eea 
This strong coupling vacuum state $\ket{0}_p\equiv\ket{0}$ will be used to construct the SU(3) magnetic basis 
in the next section.
\vspace{0.5cm} 
\begin{center} 
{\it{B. SU(3) Magnetic Basis}}
\end{center} 

\vspace{0.3cm} 

We now show that $\Sigma^\pm_{\theta_1, \theta_2}$ operating on the SU(3) plaquette magnetic basis act like a translation operators for the two gauge invariant magnetic fields. 
As shown in   Appendix    \ref{mlb}, the SU(3) magnetic basis can be written in terms of SU(3) pre-potentials \cite{manpp} as:
\begin{align}
\hspace{-0.4cm}~\ket{Z}=& \sum_{\textrm{p},\textrm{q}=0}^{\infty} \sqrt{d(\textrm{p},\textrm{q})}\frac{(a^\dagger[1] Z  b^\dagger[1])^\textrm{p}}{\textrm{ p}!}\frac{(a^\dagger[2] Z b^\dagger[2])^\textrm{q}}{\textrm{ q}!}\,\ket{0}\label{Su(3)mbasis}
\end{align}
In the above equation, the plaquette index has been suppressed and  
$$d(\textrm{p},\;\textrm{q})=\frac{1}{2} (\textrm{p}+1) (\textrm{q}+1)(\textrm{p}+\textrm{q}+2),$$
is the dimension of the  $[\textrm{p},\,\textrm{q}]$  representation of SU(3) \cite{georgi2000lie},  $Z_{\alpha\beta}$ are the elements of SU(3) matrix and correspond to the eigenvalues of $\mathcal{W}^{[{\rm p}=1,{\rm q}=1]}_{\alpha\beta}(p)$ and we have ignored plaquette index $p$ in (\ref{Su(3)mbasis}). In the axis-angle representation $Z$ 
can be written as \footnote{Advantage of this representation is that it has the following property:
\begin{align*}
&Z( \omega_1, \omega_2)Z( \theta_1, \theta_2)\\
&~~~~= e^{\left(i (\omega_1\hat{n}^{\rm a}_{[1]} + \omega_2\hat{n}^{\rm a}_{[2]}) \frac{\lambda^{\rm a}}{2}\right)}e^{\left(i (\theta_2\hat{n}^{\rm b}_{[1]} + \theta_2\hat{n}^{\rm b}_{[2]}) \frac{\lambda^{\rm b}}{2}\right)}\\
&~~~~~~~~ \because e^X e^Y= e^{X+Y+\frac{1}{2}[X,Y]+...},~~~~ [\lambda^{\rm a},\lambda^{\rm b}]=2if^{{\rm a}{\rm b}{\rm c}}\lambda^{\rm c},\\
&~~~~~~~~~~~~~~~~~~~~~~~~~~~~~f^{{\rm a b c}} \hat{n}^{\rm b}_{[h]}\hat{n}^{\rm c}_{[h']}=0,~~~~h,h'=1,2\\
&~~~~~= e^{\left(i ((\omega_1+\theta_1)\hat{n}^{\rm a}_{[1]} + (\omega_1+\theta_2)\hat{n}^{\rm a}_{[2]}) \frac{\lambda^{\rm a}}{2}\right)}\\
&~~~~~=Z(\omega_1+\theta_1,~\omega_2+\theta_2)    
\end{align*}
Which we will use to show the translation of two gauge invariant angles through the action of the disorder operator.}
\begin{align}
\!\!Z(p) = Z(\omega_1, \omega_2)
=\exp i\big(\omega_1\hat{n}^{\rm a}_{[2]} + \omega_2\hat{n}^{\rm a}_{[2]}\big) \frac{\lambda^{\rm a}}{2}.
\label{om12} 
\end{align}
In (\ref{om12}) we have  labeled the SU(3) group manifold by $Z(\omega_1, \omega_2) \equiv Z(\hat n_{[1]},\hat n_{[2]}; \omega_1, \omega_2)$. The two axes $(\hat n_{[1]}, \hat n_{[2]})$ are suppressed for the notational simplicity. 
Under SU(3) gauge transformations  at the origin (\ref{gtatorigin})   
\bea 
\ket{Z} \rightarrow \ket{\Lambda \,Z \, \Lambda^\dagger}, ~~~~\Lambda \equiv \Lambda(0,0). 
\label{su3gtzz} 
\eea 
We have used (\ref{su3gto}) and the defining equations (\ref{om12}) to obtain the above covariant transformations. The gauge transformations (\ref{su3gtzz}) show that 
\begin{align}
\omega_h  \rightarrow  \omega_h, ~~\hat n_{[h]}  \rightarrow   \Lambda ~ \hat n_{[h]} ~  \Lambda^\dagger ~~~~~~~~h=1,2. 
\label{su3gton} 
\end{align} 
Thus $(\omega_1, \omega_2)$ are the gauge invariant angles and the two axes $\hat n_{[h]} \equiv \sum_{{\rm a}=1}^8 \hat n^{\rm a}_{[h]} \lambda^{\rm a}$ transform like the adjoint vectors on every plaquette. 

In order to evaluate the action of the disorder operator  on this magnetic basis we first write down the following equations, which can be easily established using 
the commutation relations in (\ref{su3ptr})
\begin{subequations}
\begin{align}
\hspace{-0.15cm}\Sigma^+_{\theta_1, \theta_2}\, a^\dagger_{\alpha}[h] \,  \Sigma^{+\dagger}_{\theta_1, \theta_2}=&  \left(a^\dagger[h] ~ e^{i\left(\theta_1\hat{n}^{\rm a}_{[2]} + \theta_2\hat{n}^{\rm a}_{[2]} \right)\frac{\lambda^{\rm a}}{2}}\right)_{\alpha},\\
\hspace{-0.15cm}\Sigma^-_{\theta_1, \theta_2}\, b^{\dagger}_{\alpha}[h] \,  \Sigma^{-\dagger}_{\theta_1, \theta_2}= & \left(  e^{-i\left(\theta_1\hat{n}^{\rm a}_{[1]} + \theta_2\hat{n}^{\rm a}_{[2]} \right)\frac{\lambda^{\rm a}}{2}}~b^{\dagger}[h]\right)_{\alpha}  
\end{align}
\end{subequations}
Using the above equations we can easily prove that 
\begin{align*}
\Sigma^+_{\theta_1, \theta_2}    \ket{Z(\omega_1,\omega_2)}= &\ket{e^{\left(i (\theta_1\hat{n}^{\rm a}_{[1]}+ \theta_2\hat{n}^{\rm a}_{[2]}) \frac{\lambda^{\rm a}}{2}\right)}Z(\omega_1,\omega_2)}\\
=& \ket{Z(\omega_1+\theta_1,\omega_2+\theta_2)}, \\
\Sigma^-_{\theta_1, \theta_2}    \ket{Z(\omega_1,\omega_2)}= &\ket{Z(\omega_1,\omega_2)e^{\left(-i (\theta_1\hat{n}^{\rm a}_{[1]} + \theta_2\hat{n}^{\rm a}_{[2]}) \frac{\lambda^{\rm a}}{2}\right)}}\\
=& \ket{Z(\omega_1-\theta_1,\omega_2-\theta_2)} 
\end{align*}
or,
\begin{align}
   \Sigma^\pm_{\theta_1, \theta_2}\;\ket{Z(\omega_1, \omega_2)}=  \;\ket{Z(\omega_1\pm\theta_1,\; \omega_2\pm\theta_2)} 
\end{align}
Therefore the disorder operator in (\ref{SU(3)do}) translate two gauge invariant angles. We can thus interpret them as creation-annihilation operators for SU(3) magnetic vortices.
\newpage
\vspace{0.5cm} 
\begin{center} 
{{\it C. ~SU(3) Order-Disorder Algebra}}
\end{center} 
\vspace{0.3cm}

The  SU(3) order-disorder algebra is 
\begin{align}
\begin{aligned} 
\Sigma^+_{[\vec \theta]}(p) \;\mathcal{W}^{[\textrm{p},\,\textrm{q}]}_{\alpha\beta}(p)\;\Sigma^{+\dagger}_{[\vec \theta]}(p)\; & = & \; D^{[\textrm{p},\,\textrm{q}]}_{\alpha\gamma}(\vec \theta)\, \mathcal{W}^{[\textrm{p},\,\textrm{q}]}_{\gamma\beta}(p) \\
\Sigma^-_{[\vec \theta]}(p) \mathcal{W}^{[\textrm{p},\,\textrm{q}]}_{\alpha\beta}(p)\Sigma^{-\dagger}_{[\vec \theta]}(p)\; & = & \; \mathcal{W}^{[\textrm{p},\,\textrm{q}]}_{\alpha\gamma}(p) \,D^{[\textrm{p},\,\textrm{q}]}_{\gamma\beta}(\vec \theta)   
\label{su3:order-disorder}
\end{aligned} 
\end{align}
In (\ref{su3:order-disorder}), $D^{[\textrm{p},\,\textrm{q}]}(\theta_1,\theta_2) \equiv \exp\left(i (\theta_1\hat{n}^{\rm a}_{[1]} + \theta_2\hat{n}^{\rm a}_{[2]} )  \frac{\lambda^{\rm a}}{2}\right)$ is SU(3) Wigner D-matrix in the $[\textrm{p},\,\textrm{q}]$ representation. Like is SU(2) case, we have used the dual canonical commutation relations (\ref{dccr}) to obtain the SU(3) order-disorder algebra
in (\ref{su3:order-disorder}).

\subsubsection[3]{\bf SU(N) Disorder Operator}
\label{sundisop}
We now use the SU(N) dual electric scalar potentials ${\cal E}(p)$ in (\ref{edt})  to define the SU(N) disorder operator 
 \begin{equation}
 \Sigma^\pm_{[\theta_1\theta_2\cdots \theta_{\text{N}-1}]}(p)= \exp\;i \left\{\; \vec \theta(p) \cdot \; \vec {\cal E}_\pm(p)\right\}.   
 \label{mdo} 
 \end{equation}
In (\ref{mdo}) $[\theta_1,\theta_2,\cdots , \theta_{\text{N}-1}] \; \equiv \; [\theta_1(p),\theta_2(p),\cdots , \theta_{\text{N}-1}(p)]$ are the $(\text{N}-1)$ external angular parameters characterizing the SU(N) disorder operator on the plaquette $(p)$ and 
\bea 
\vec \theta(p) \equiv \sum_{h=1}^{(\text{N}-1)} \theta_h(p) \hat n_{[h]}(p). \label{axistheta}
\eea 
The  invariance (\ref{gtatorigin}) demands that the  operator $\vec{\theta}(p)$ in (\ref{mdo}) is the most general vector operator constructed out of magnetic flux operator ${\cal W}_{\alpha\beta}(p)$. In other words,  they depend on the $(\text{N}-1)$ directions of the SU(N) magnetic fields. 
 In SU(2) and SU(3) cases in the previous sections  we have already constructed  one and two independent 
 axes respectively using the plaquette magnetic flux operators. In the same way  we now iteratively define the $(\text{N}-1)$ linearly independent ``SU(N) magnetic 
axes" using the SU(N) symmetric structure constants $d^{{\rm a}{\rm b}{\rm c}}$  as follows:
\begin{equation} 
\vec {n}^{\rm a}_{[h+1]}(p) \equiv d^{{\rm a}{\rm b}{\rm c}}\; \vec {n}^{b}_{[h]}(p)\; \vec {n}^{c}_{[1]}(p)~\;h=1,2,\dots, \text{N}-2. 
\end{equation} 
\noindent 
The first magnetic axis is defined as $ \vec {n}^{\rm a}_{[1]}(p) \equiv  \text{Tr} (\Lambda^{\rm a} \left( {\cal W}+{\cal W}^\dagger\right))$ where $\Lambda^{\rm a} ({\rm a}=1,2,\cdots,(\text{N}^2-1))$ are the SU(N) 
fundamental representation matrices. The iterative procedure ends as \footnote{We have used the property:
$(d^{{\rm ab e}}d^{{\rm cd e}}+d^{{\rm ac e}}d^{{\rm bd e}}+d^{{\rm ade}}d^{{\rm bc e}})=\tfrac{1}{3}(\delta ^{{\rm a  b}}\delta^{{\rm c d}}+\delta ^{{\rm a c}}\delta^{{\rm b d}}+\delta ^{{\rm b c}}\delta^{{\rm a d}})$ for SU(3). It can be similarly 
generalized to SU(N) with $(\text{N}-1)$ d structure functions.} 
$\vec n^{\rm a}_{[\text{N}]} \equiv 
d^{{\rm a}{\rm b}{\rm c}}\; \vec {n}^{{\rm b}}_{[\text{N}-1]}(p)\; \vec {n}^{\rm c}_{[1]}(p) = \vec n^{\rm a}_{[1]}(p)$.  The $(\text{N}-1)$ SU(N) magnetic field operators $
\vec n^{\rm a}_{[h]}; ~h=1,2,\cdots,(\text{N}-1)$ are Hermitian as the symmetric structure constants $d^{{\rm a}{\rm b}{\rm c}}$ are always real.   Under gauge transformation (\ref{gtatorigin}), these axes transform as vectors
\begin{equation}
 \vec {n}^{\rm a}_{[h]}(p)\rightarrow ~ R^{\rm ab}(\Lambda)\;\vec{n}^{\rm b}_{[h]}(p) ,~~ h=1,2,\cdots, (\text{N}-1).
\end{equation}
The  disorder operator is invariant  under the gauge transformations (\ref{gtatorigin}) as $\vec \theta(p)$ and  the dual electric potentials  $\vec {\cal E}(p)$
both transform as vectors. As in the case of SU(2) (see (\ref{ssdi})) and SU(3) (see (\ref{ssdi3})),
 $ \Sigma^{+}_{[\vec \theta]}(p)$ 
and $  \Sigma^{-}_{[\vec \theta]}(p)$ are not independent and satisfy
\begin{align}
\begin{split}
 \Sigma^{+}_{[\vec \theta]}(p) \; \Sigma^{-}_{[\vec \theta]}(p)\; = \; \mathcal{I}\; =\; \Sigma^{-}_{[\vec \theta]}(p)\;  \Sigma^{+}_{[\vec \theta]}(p). \;\;
\end{split}
\label{invdo} 
\end{align} 
Here $\mathcal{I}$ is unity operator in the physical Hilbert space.
The relations (\ref{invdo}) follow from the parallel 
transport relating the two electric scalar potentials:
$ \mathcal{E}^{\rm a}_-(p)=-R^{\rm ab}( \mathcal{W}(p))\;  \mathcal{E}^{\rm b}_+(p)$ and ${\hat n}^{\rm a}_{[h]}(p)=-R^{\rm ab}( \mathcal{W}(p)) \; {\hat n}^{\rm b}_{[h]}(p); ~~h =1,2\cdots, (\text{N}-1)$.
We now briefly disscuss the SU(N) prepotential operators to be used in the  Section \ref{sundisop}\;-\;B for  the construction of SU(N) magnetic basis. 
\vspace{0.4cm}

\begin{center} 
{\it{A. SU(N) Prepotential Operators}}
\end{center}
\vspace{0.2cm} 

\noindent The SU(N) dual electric scalar potentials ${\cal E}^a(p)$ can be written in terms of the $(\text{N}-1)$ N-plets  of harmonic oscillators at each of the two ends of the plaquette $p$. We define   
\begin{align}
\begin{aligned} 
\mathcal{E}_+^{\mathrm a}(p) =&~~\, \frac{1}{2} \sum_{h=1}^{(\text{N}-1)}~ \left[
\sum_{\alpha,\,\beta=1}^\text{N} a^{\dagger}_\alpha[h]\;(\Lambda^{\mathrm a})_{\alpha\beta} \, a_\beta[h]\right], \\
\mathcal{E}_-^{\mathrm a}(p)=&\, -\frac{1}{2}
\sum_{h=1}^{(\text{N}-1)}\left[ \sum_{\alpha,\,\beta=1}^\text{N} b_\alpha[h]\;(\Lambda^{\mathrm a})_{\alpha\beta} \, b^\dagger_\beta[h]\right].
\label{sunppp} 
 \end{aligned} 
\end{align}
In (\ref{sunppp}), we have introduced prepotential N-plets $(a_\alpha[h], a^\dagger_\alpha[h])$ 
and $(b_\alpha[h], b^\dagger_\alpha[h])$  for each of the $(\text{N}-1)$  fundamental  representations 
of SU(N). They are denoted by $h=1,2,\cdots,(\text{N}-1)$ and 
we have suppressed the additional plaquette index on the right hand side of (\ref{sunppp}) for 
convenience. The  $\frac{\Lambda^a}{2}$ are the $(\text{N}^2-1)$ SU(N) matrices in the fundamental representation. The harmonic oscillator commutation relations of the SU(N) prepotentials  imply
 \begin{align}
\begin{aligned}
 \left[\mathcal{E}_+^{\mathrm a}[h],~a^{\dagger}_\alpha[h']\right] ~=~&\;~~  \delta_{h,h'}\; \frac{1}{2}\, a^{\dagger}_\beta[h] \;\Lambda^{\mathrm a}_{\beta\alpha},\\
 \left[\mathcal{E}_+^{\mathrm a}[h],~ b^{\dagger}_\alpha[h']\right] ~=~&- \delta_{h,h'}\;\frac{1}{2} \;\Lambda^{\mathrm a}_{\alpha\beta} \;b^{\dagger}_\beta[h].
 \label{sunepp}
 \end{aligned}
 \end{align}
We also note that under SU(N) gauge transformations (\ref{gtatorigin}) with $\Lambda \equiv \Lambda(0,0)$ (see Figure \ref{DT}-b) these oscillators transform as
\begin{align}
\begin{aligned}
    a^\dagger_\alpha[h]\rightarrow & ~~   a^\dagger_\beta[h]~ \Lambda_{\beta\alpha}\; , 
    ~~~~~ \forall\; h=1,2,...,(\text{N}-1),\\
 b^\dagger_\alpha[h]\rightarrow & ~~   \Lambda^\dagger_{\alpha\beta}\; b^\dagger_\beta[h]\; ,
 ~~~~~ \forall\; h=1,2,...,(\text{N}-1).
 \label{sunppt}
 \end{aligned} 
\end{align}
Like in SU(2) and SU(3) cases, the relations (\ref{sunepp}) 
and (\ref{sunppt}) will be useful in constructing the SU(N)
magnetic basis in the next section.
\newpage
\vspace{0.5cm} 
\begin{center} 
{\it{B. SU(N) Magnetic Basis}}
\end{center} 
\vspace{0.3cm} 
In this section, we construct the  SU(N) magnetic basis for all SU(N) and show that the disorder operators on a magnetic basis act as  shift operators for the N-1 magnetic fields. The SU(N) magnetic basis has been constructed in Appendix \ref{mlb} and is given by:
\begin{equation}
    \ket{ Z}=\sum_{[\vec j]=0}^\infty \sqrt{d(\vec j)} \prod_{h=1}^{\text{N}-1} \frac{1}{j_h!}\left( a^
\dagger[h] Z b^\dagger[h]\right)^{2j_h}\; \ket{0}.\label{sunmb}
\end{equation}
In (\ref{sunmb}) $\sqrt{d(\vec j)}$ is the dimension of the SU(N)  $[\vec{j}]\, (\equiv (j_1,j_2,\cdots,j_{\text{N}-1})$ representation.  The SU(N) strong coupling vacuum $\ket{0}$ in the dual description on every plaquette satisfies
\bea
{a}_\alpha[h] \;\ket{0} =0,~{b}_\alpha[h] \;\ket{0} =0, ~~h=1,2, \cdots ,(\text{N}-1).
\label{scvsn}
\eea
Like in SU(2) and SU(3) cases we  parameterize the SU(N) matrix $Z\equiv Z(p)$  in (\ref{sunmb}) on every plaquette $p$ in the angle-axis representation as 
\begin{equation}
Z=  Z(\omega_1,\omega_2, \cdots, \omega_{\text{N}-1}) = \exp{ i \left(\omega_{h} \,\hat{n}^{\rm a}_{[h]}\,\frac{\Lambda^{\rm a}}{2}\right)}. \label{Zsun}
\end{equation}
\noindent In (\ref{Zsun})  the $(\text{N}-1)$ linearly independent unit vectors are defined as
\begin{equation}
\vec{n}^{\rm a}_{[r+1]}(p) \equiv \;d^{{\rm a}{\rm b}{\rm c}}\; \hat{n}^{\rm b}_{[r]}(p)\; \hat{n}^{\rm c}_{[1]}(p),~\;r=1,2,\dots, \text{N}-2.  
\end{equation}
We have again suppressed the $(\text{N}-1)$ axes $\hat n^a_{[h]}$
in $Z(\omega_1,\omega_2, \cdots, \omega_{\text{N}-1})$ for the notational simplicity. 

In order to evaluate the action of disorder operators on the magnetic basis (\ref{sunmb}), we use (\ref{sunepp}) to obtain:
\begin{align}
 \Sigma^+_{[\vec\theta]} \;a^\dagger_{\alpha}[h] \;  \Sigma^{+\dagger}_{[\vec\theta]}=&  \left(a^\dagger[h] \; e^{i\left(\theta_h\hat{n}^{\rm a}_{[h]} \right)\frac{\Lambda^{\rm a}}{2}}\right)_{\alpha},\\
 \Sigma^-_{[\vec\theta]} \; b^{\dagger}_{\alpha}[h] \;  \Sigma^{-\dagger}_{[\vec\theta]}= & \left(e^{-i\left(\theta_h\hat{n}^{\rm a}_{[h]} \right)\frac{\Lambda^{\rm a}}{2}}\; b^{\dagger}[h]\right)_{\alpha}   
\end{align}
Therefore the action of disorder operators on the magnetic basis is given by 
\begin{align}
\Sigma^+_{[\vec\theta]}\;\ket{ Z}=&\;\ket{ e^{i\left(\theta_h\hat{n}^{\rm a}_{[h]}\right)\frac{\Lambda^{\rm a}}{2}}~Z}\\
\Sigma^-_{[\vec\theta]}\;\ket{ Z}=&\;\ket{ Z~e^{i\left(\theta_h\hat{n}^{\rm a}_{[h]}\right)\frac{\Lambda^{\rm a}}{2}}}
\end{align}
\usetikzlibrary{shapes.misc}
\tikzset{cross/.style={cross out, draw=black, minimum size=2*(#1-\pgflinewidth), inner sep=0pt, outer sep=0pt},
cross/.default={1.5pt}}
\tikzset{%
  dots/.style args={#1per #2}{%
    line cap=round,
    dash pattern=on 0 off #2/#1
  }
}
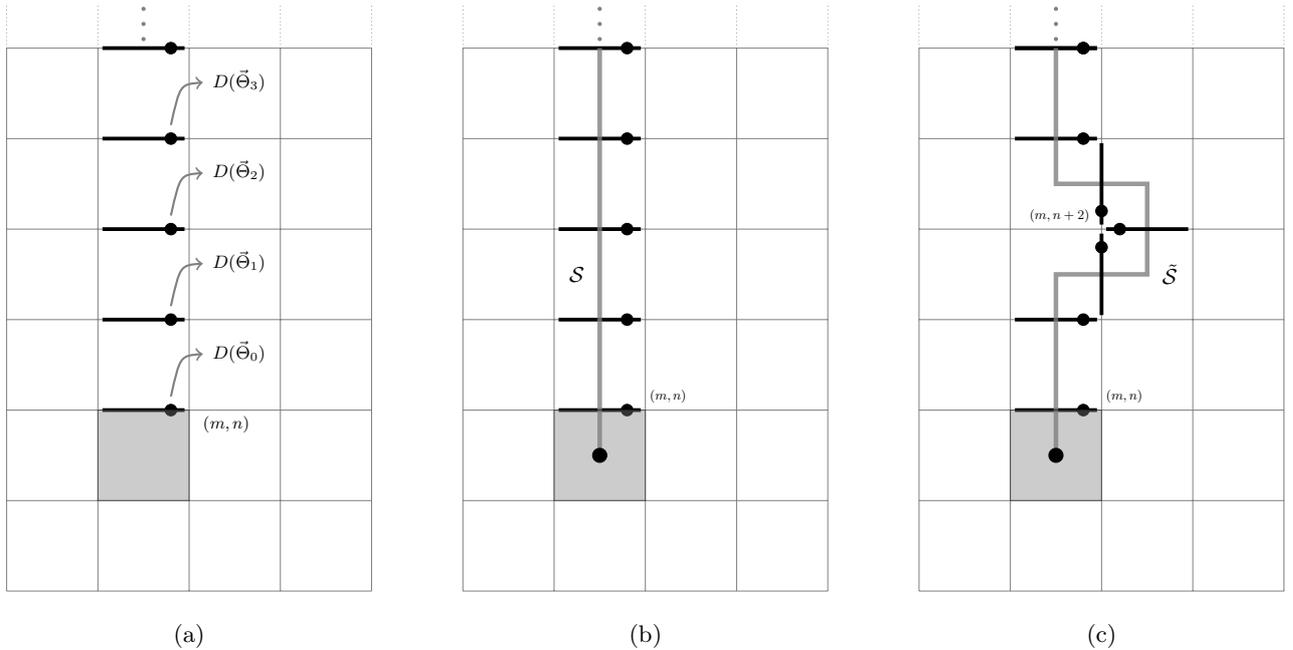
\begin{figure*}[t]
\begin{tikzpicture}[scale=1.2]
\draw[help lines,gray] (0,0) grid (4,6);
\foreach [count=\i, evaluate=\i as \x using int(\i+1)] \y in {0,1,2,3,4}{
\draw [gray,dots=20 per 1cm] (\y,6)--(\y,6.5);
}
\begin{scope}[shift={(1,1)}]
\foreach \y in {0,1,2,3}{
\draw [line width=.5mm] (.05,\y+1)--(.95,\y+1);
\draw [fill] (.8,\y+1) circle (1.8pt);
\draw[gray, thick,->] (.8,\y+1.15).. controls (.90, \y+1.6)..(1.15, \y+1.62);
\node[scale=.75] at ( 1.55, \y+1.64) {$D(\vec{\Theta}_{\y})$}; 
}
\draw [gray,ultra thick,dots=5 per 1cm] (0.5,5.1)--(0.5,5.5);
\draw [line width=.5mm] (.05,5)--(.95,5);
\draw [fill] (.8,5) circle (1.8pt);
\node [scale=.7,below] at (1.41,1.) {$(m,n)$};
\draw [fill=gray,opacity=.4] (0,0) rectangle (1,1);
\end{scope}
\node[] at (2.,-.5) {(a)};
\begin{scope}[shift={(5,0)}]
\draw[help lines,gray] (0,0) grid (4,6);
\foreach [count=\i, evaluate=\i as \x using int(\i+1)] \y in {0,1,2,3,4}{
\draw [gray,dots=20 per 1cm] (\y,6)--(\y,6.5);
}
\begin{scope}[shift={(1,1)}]
\foreach \y in {1,2,3,4}{
\draw [line width=.5mm] (.05,\y)--(.95,\y);
\draw [fill] (.8,\y) circle (1.8pt);
}
\draw [line width=.5mm] (.05,5)--(.95,5);
\draw [fill] (.8,5) circle (1.8pt);
\draw [gray,line width=.6mm,opacity=.8] (.5,.5)--(.5,5);
\draw [gray,ultra thick,dots=5 per 1cm] (.5,5.1)--(.5,5.5);
\draw [fill=gray,opacity=.4] (0,0) rectangle (1,1);
\node[ scale=.9] at ( .25,2.5) { $\mathcal{S}$};
\node[ scale=.55] at ( 1.25,1.15) { $(m,n)$};
\draw [fill] (.5,.5) circle (2.2pt);
\end{scope}
\node[] at (2.,-.5){(b)};
\end{scope}
\begin{scope}[shift={(10,0)}]
\draw[help lines] (0,0) grid (4,6);
\foreach [count=\i, evaluate=\i as \x using int(\i+1)] \y in {0,1,2,3,4}{
\draw [gray,dots=20 per 1cm] (\y,6)--(\y,6.5);
}
\begin{scope}[shift={(1,1)}]
\foreach \y in {1,2,4,5}{
\draw [line width=.5mm] (.05,\y)--(.95,\y);
\draw [fill] (.8,\y) circle (1.8pt);
}
\draw [line width=.5mm] (.05,5)--(.95,5);
\draw [fill] (.8,5) circle (1.8pt);
\draw [gray,line width=.6mm,opacity=.8] (.5,.5)--(.5,2.5)--(1.5,2.5)--(1.5,3.5)--(.5,3.5)--(.5,5);
\draw [gray,ultra thick,dots=5 per 1cm] (.5,5.1)--(.5,5.5);
\draw [fill=gray,opacity=.4] (0,0) rectangle (1,1);
\node[ scale=.9] at ( 1.75,2.5) { $\tilde{\mathcal{S}}$};
\node[ scale=.55] at ( 1.25,1.15) { $(m,n)$};
\node[ scale=.55] at ( .54,3.15) { $(m,n+2)$};
\draw [fill] (.5,.5) circle (2.2pt);
\draw [line width=.5mm] (1.05,3)--(1.95,3);
\draw [fill] (1.2,3) circle (1.8pt);
\draw [line width=.5mm] (1.,3.05)--(1,3.95);
\draw [fill] (1,3.2) circle (1.8pt);
\draw [line width=.5mm] (1.,2.05)--(1,2.95);
\draw [fill] (1,2.8) circle (1.8pt);
\end{scope}
\node[] at (2,-.5) {(c)};
\end{scope}
\end{tikzpicture}
  \caption{ (a) The disorder operator $\Sigma^+_{[\vec{\theta}]}(m,n)$, defined in equation (\ref{disks}) rotates all horizontal links $U(m-1,n';\hat{1}), \, \forall \; n'\geq n$ around an axis $ \vec{\Theta}(m,n')$ (for $n'= n, n+1, n+2,..$ they are denoted by  $\vec{\Theta}_0,\vec{\Theta}_1,\vec{\Theta}_2,...$), (b) Invisible SU(N) Dirac string ${\cal S}$. The rotated links ${\sf l} \in {\cal S}$ are the dark horizontal links, (c) Shape of Dirac string can be deformed without affecting the endpoint or the location of the magnetic vortex. The SU(N) gauge transformations at site $(m,n+2)$ changes the shape of the Dirac string from ${\cal S}$ to $\tilde{\cal S}$.}\label{dstring}
\end{figure*}
We now use  axis angle-representation (\ref{Zsun}) to get
\begin{align}
\begin{aligned}
\Sigma^+_{[\vec{\theta}]}\;\ket{ Z(\omega_{h})}=&\;\ket{ e^{i \left(\theta_{h} \hat{n}^{\rm a}_{[h]}\right)\frac{\Lambda^{\rm a}}{2}}e^{i \left(\omega_{h} \hat{n}^{\rm a}_{[h]}\right)\frac{\Lambda^{\rm a}}{2}}}
\\
=&\;\ket{Z(\omega_{h}+\theta_{h})}\\
\Sigma^-_{[\vec{\theta}]}\;\ket{ Z(\omega_{h})}=&\;\ket{ e^{i \left(\omega_{h} \hat{n}^{\rm a}_{[h]}\,\right)\frac{\Lambda^{\rm a}}{2}}e^{-i \left(\theta_{h} \hat{n}^{\rm a}_{[h]}\,\right)\frac{\Lambda^{\rm a}}{2}}}\\
=&\;\ket{Z(\omega_{h}-\theta_{h})}
\end{aligned}\label{addition}
\end{align}
Therefore the disorder operator on a plaquette $p$ translates the $\text{N}-1$ gauge invariant angles defining the SU(N) magnetic fluxes.

\vspace{0.3cm} 
\begin{center} 
{{\it C. ~SU(N) Order-Disorder Algebra}}
\end{center} 
\vspace{0.3cm}

Using the canonical commutation relations in the dual description (\ref{dccr}) we get 
 Similarly, the SU(N) order-disorder algebra is
\begin{align} 
\begin{aligned} 
\Sigma^+_{[\vec \theta]} (p)\;\mathcal{W}^{[\vec j]}_{\alpha\beta}(p) \;\Sigma^-_{[\vec \theta]} (p)= D^{[\vec j]}_{\alpha\gamma}([\vec \theta]) \; \mathcal{W}^{[\vec j]}_{\gamma\beta}(p)\\
\Sigma^-_{[\vec \theta]} (p)\;\mathcal{W}^{[\vec j]}_{\alpha\beta}(p) \;\Sigma^+_{[\vec \theta]} (p)=  \; \mathcal{W}^{[\vec j]}_{\alpha\gamma}(p) D^{[\vec j]}_{\gamma\beta}([\vec \theta]).
\label{sunoda1} 
\end{aligned} 
\end{align}
In (\ref{sunoda1}) the Wigner matrix $D^{[\vec j]}([\vec \theta])$ represent the SU(N) rotations around the  magnetic axes $\hat n_{[h]}$ by $\theta_{h}$ with $h=1,2,\cdots ,(\text{N}-1)$. 

\subsubsection*{Reduction to 't Hooft Algebra} In the special case when the rotations are in 
the center of SU(N)  with 
$Z \in Z_\text{N}$ and $Z^\text{N}=1$, 
we get $$D^{[\vec j]}(Z) = (z)^{\eta[\vec j]}~\;{\cal I}, ~~~ z^\text{N}=1 $$ 
where ${\cal I}$ is $\text{N}\times \text{N}$ unit matrix and $\eta[\vec j]$ 
is the N-ality of the  $[\vec j]$ representation.  
We thus get the  't Hooft Wilson order-disorder algebra \cite{hooft,tomb,dis_nonabelian,mack}.
\begin{align} 
\begin{aligned} 
\Sigma^+_{[Z_\text{N}]} (p)\;\mathcal{W}^{[\vec j]}_{\alpha\beta}(p) \;\Sigma^-_{[Z_\text{N}]} (p) ~& = &  (z)^{\eta[\vec j]} \; \mathcal{W}^{[\vec j]}_{\alpha\beta}(p)\\
\Sigma^-_{[Z_\text{N}]} (p)\;\mathcal{W}^{[\vec j]}_{\alpha\beta}(p) \;\Sigma^+_{[Z_\text{N}]} (p) ~& = & (z)^{\eta[\vec j]} \; \mathcal{W}^{[\vec j]}_{\alpha\beta}(p).
\label{sunoda1z} 
\end{aligned} 
\end{align}
The SU(N) center elements in (\ref{sunoda1z}) are $$z = e^{\frac{2\pi i m}{\text{N}}}, ~m=0,1,\cdots, (\text{N}-1).$$ 



\section{SU(N) Dirac Strings}\label{ids}
The disorder operators defined in the previous section can also be written in terms of the Kogut-Susskind link holonomies and their electric fields using the exact duality transformations (\ref{edt}). As expected,  these Disorder operators $\Sigma(p)$ are highly non-local operators in the original description but their physical action is essentially  local.
 Using duality transformation relation (\ref{edt}) we write;
  \begin{align}
\begin{aligned}\label{disks}
&~~~\Sigma^+_{[\theta_1\theta_2\cdots \theta_{\text{N}-1}]}(m,n)\\
&= \exp \left\{ i \; \vec\theta^{\rm a}(m,n)\cdot\!\!\! \sum_{n'=n+1}^\infty R^{\rm ab} ({\sf S}(m,n;n')) E^{\rm b}_-(m,n';\hat{1})\right\}
\end{aligned}
\end{align}
In (\ref{disks}) the parallel transports \footnote{Here we have redefined the parallel transport ${\sf S}(m,n;n')\equiv {\sf T}^\dagger(m-1,n)S(m,n;n')$ and magnetic axis $\vec{n}^{\rm a}_{[1]}(m,n)\equiv R^{\rm ab}({\sf T} (m-1,n))\vec{n}^{\rm b}_{[1]}(p)= \text{Tr} ( \Lambda^{\rm a}( U_p(m,n) +U^\dagger_p(m,n)))$. The advantage of using new parallel transports ${\sf S}(m,n;n')$ is that they are not connected to the origin and therefore  more appropriate for the original Kogut-Susskind  formulation.} are given by $$ {\sf S}(m,n;n')=\;U(m-1,n;\hat{1})\;\prod_{q=n}^{n'} U(m,q,\hat{2})$$ and the axis of rotation

\begin{equation}
\vec \theta(m,n)= \sum_{h=1}^{\text{N}-1} \theta_h\hat{n}_{[h]}(m,n)
\end{equation} 
is characterized by $\text{N}-1$ external angular parameters $[\theta_1, \theta_2,...,\theta_{\text{N}-1}]=[\theta_1(m,n), \theta_2(m,n),...,\theta_{\text{N}-1}(m,n)]$.  The  $\text{N}-1$  linearly independent magnetic axes are defined as
\begin{align} 
\vec {n}^{\rm a}_{[r+1]}(m,n) \equiv&\; d^{{\rm a}{\rm b}{\rm c}}\; \vec {n}^{b}_{[r]}(m,n)\; \vec {n}^{c}_{[1]}(m,n),\\
~~~~~~~~~~~~~~~& \forall\;r=1,2,\dots, (\text{N}-2). \nonumber
\end{align} 
The first magnetic axis is defined as  $\vec{n}^{\rm a}_{[1]}(m,n)= \text{Tr} ( \Lambda^{\rm a}( U_p(m,n) +U^\dagger_p(m,n)))$. This operator rotate all the horizontal link $U(m-1,n';\hat{1}),~n'\geq n$. We can define the axis of rotation associated with each rotated link as $\vec{\Theta}^{\rm a}(m, n'>n)= R^{\rm ab} ({\sf S}(m,n; n'))\, \vec{\theta}^{\;\rm b}(m,n)$ which can also we recast in an iterative relation 
\begin{equation}\label{theta}
\vec{\Theta}^{\rm a}(m,n'+1)= R^{\rm ab} (U(m,n';\hat{2}))\;\vec{\Theta}^{\rm b}(m,n')
\end{equation} 
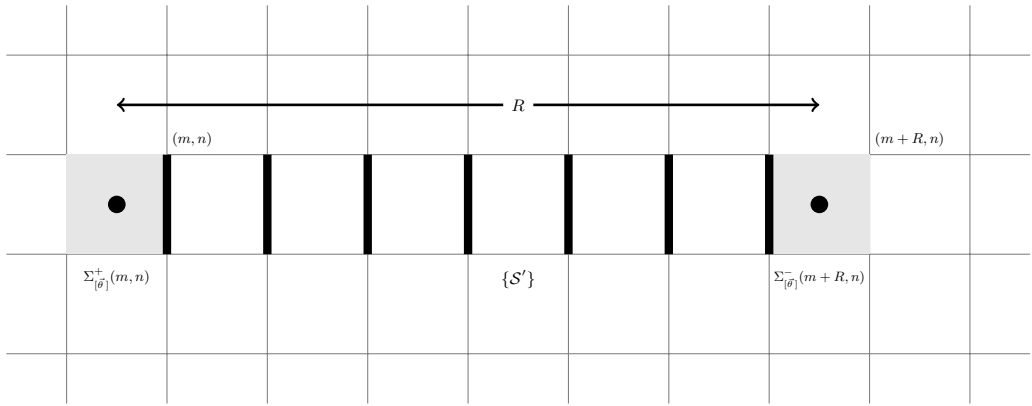
\begin{figure*}[t]
\begin{center}
\begin{tikzpicture}[scale=1.32]
\draw[ gray] (.4,-.5) grid (10.6,3.5);

\fill[gray!20] (1,1) rectangle (2,2);
\fill[gray!20] (8,1) rectangle (9,2);
\foreach \x in {2,3,4,...,7,8}
{\draw[line width=1.1mm]  (\x,1)-- (\x,2);
}
\node[ scale=.6] at ( 2.25, 2.15){$(m,n)$};
\node[ scale=.6] at ( 9.4, 2.15){$(m+R,n)$};
\node[ scale=.7] at ( 5.5, 2.5){$R$};
\node[ scale=.7] at ( 5.5, .75){${\{\cal S}'\}$};
\node[ scale=.6] at ( 1.5, .75){$\Sigma^+_{[\vec \theta\,]}(m,n) $};
\node[ scale=.57] at ( 8.5, .75){$\Sigma^-_{[\vec \theta\,]}(m+R,n)$};
\draw[line width=.35mm,<-] (1.5, 2.5)--(5.35,2.5);
\draw[line width=.35mm,<-] (8.5, 2.5)--(5.65,2.5);
\fill [] (1.5,1.5) circle (2.5 pt);
\fill [] (8.5,1.5) circle (2.5 pt);
\end{tikzpicture}
\end{center}
\caption{Action of the disorder operators $\Sigma^+_{[\vec \theta]}(m,n) \Sigma^-_{[\vec \theta]}(m+R,n)$ 
creating SU(N) vortex-antivortex at a distance $R$ apart. The SU(N) transformations rotate the dark vertical links denoted by ${\sf l}' \in {\cal S}'$ in (\ref{ssdcf}). This set of vertical dark links ${\sf l}'$ is denoted by ${\{\cal{S}'}\}$.}\label{vortexpair} 
\end{figure*}

 We can similarly obtain $\Sigma^{-}_{[\vec \theta\,]}$ by using (\ref{edt}), (\ref{lrreld}) and (\ref{mdo}). Now we have 
  \begin{equation}\label{diskss}
\Sigma^+_{[\vec\theta]}(m,n)= \exp \left\{ i \!\!\! \sum_{n'=n+1}^\infty \!\! \vec\Theta^{\rm a} (m,n')\cdot E^{\rm a}_-(m,n';\hat{1})\right\}
\end{equation}
They rotate the links 
  \bea \label{diskss2}
\Sigma^+_{[\vec{\theta}]}(m,n)U_{\alpha\beta}(m,n';\hat{1})\Sigma^{+\dagger}_{[\vec{\theta}]}(m,n) \phantom{xxxxxxxx} ~~~~~~~~ ~~~~~~\\
= \, U_{\alpha\gamma}(m,n';\hat{1})D_{\gamma\beta}( \vec{\Theta}(m,n')),~~~\forall \;n' \;\geq \;n. ~~~~ \nonumber
\eea
These rotations of the horizontal link holonomies are shown in figure \ref{dstring}-a. The rotational 
axes 
of these link holonomies are related through the parallel transport equations (\ref{theta}) which, in turn, are  obtained by the exact duality transformations (\ref{edt}). These special relations ensure that 
they create magnetic flux only on the plaquette located at the end point $(m,n)$ keeping all the other plaquette fluxes unaffected (see Appendix \ref{app:invisibility}). Therefore this local action by the  non-local operator (\ref{diskss}) creates  an invisible non-abelian Dirac string ${\cal S}$ originating from the corresponding plaquette (see Figure \ref{dstring}-b). In Appendix \ref{app:invisibility} is shown that using gauge transformations these Dirac strings can be deformed arbitrarily except their gauge invariant endpoints.

\section{Path Integral Representation}\label{pathrep} 


In this section, we construct the  path integral representation of the SU(N) disorder operators so that their behaviour can also be studied 
using Monte-Carlo simulations in future studies. Such construction for the $Z_2$ 't Hooft 
disorder operator in pure SU(2) lattice gauge theory can be found in \cite{mack,dis_nonabelian}. 
\noindent The ground state wave wave functional  depends on the 
links in the 2-dimensional surface $\Sigma$ at time $t=0$ \cite{mack}:
\begin{equation}
 \Psi_0(U) \equiv \langle U |  {\psi}(0) \rangle =\!\!\int \prod_{l>0}  dU(l)
 e^{
 \beta \sum\limits_{p >0} \text{Tr}(U_p+U^\dagger_p)}\label{pffgs}
\end{equation}
\noindent In (\ref{pffgs}) the integration is done over all links $l >0$ which are the links at time $t > 0$. Similarly the plaquettes involved in the summation are in the upper half lattice at $t > 0$.  Thus the ground state $\Psi_0(U)$ depends only on the spatial links at $t=0$. The  expectation values of any functional $F[U(l)]$ in the ground state $\ket{\psi(0)}$ is defined as $$\langle F[U(l)]\rangle= \bra{\psi(0)} F[U(l)]\ket{\psi(0)}.$$
The path integral representation is 
\begin{equation}\label{expectF}
\langle F[U(l)]\rangle =\frac{1}{Z(\beta)} \int d\mu(U)
F[U(l)]~ e^{\beta\text{tr} \left(U_p+U_p^\dagger)\right)},
\end{equation}
where $d\mu(U) \equiv \prod_{l}\; dU(l)$ and $l, p$ now denote all the links and plaquettes in the 3-dimensional lattice and $\beta= \frac{2\text{N}}{g^2}$. 
The partition function $Z(\beta)$  is given by:
\begin{align}
Z(\beta) = \int\prod_{l} d U(l) e^{\beta\sum_{p}(\text{tr} (U_p+U_p^\dagger))}
\end{align} 


\noindent The action of $\Sigma^+_{[\vec \theta]} (m,n)$ rotates  all the links crossing the Dirac string 
by the appropriate  SU(N) Wigner D matrices as shown in Figure \ref{dstring}-a. Therefore the expectation value  $\Sigma^+_{[\vec \theta]}(m,n)$ or the free energy 
of the SU(N)  
magnetic vortex can be defined as
\begin{align}
\langle \,\Sigma^+_{[\vec \theta]} (m,n)\,\rangle   
&=\left\langle   e^{-\beta\sum\limits_{{\sf l}\in {\cal S}}\left[\text{tr}(D(\vec \theta)U_p+U_p^\dagger D^\dagger(\vec \theta))-\text{tr}(U_p+U_p^\dagger) \right]}    \right\rangle    \nonumber  \\
& \equiv ~~~~ e^{-\beta  F_{mag}(\;\vec \theta\;)}.
\label{dosim}
\end{align}
In (\ref{dosim})  the summation sign  includes only those plaquettes which protrude from  the links  ${\sf l} \in \{{\cal S}\}$  (see Figure \ref{dstring}-b) in the $+$ve time direction and $F_{mag}(\;\vec \theta\;)$ denotes the free energy of the magnetic vortex. 
Note that the path-integral representation 
 for the SU(N) vortex (\ref{dosim}) is   analogous to the path integral representations for the  defects in 2-d Ising model \cite{kada}   and 
 $Z_\text{N}$ vortices in SU(N) gauge theory \cite{hooft} obtained by Kadanoff 
 and 't Hooft  respectively. 
 We can also define SU(N)
electric free energy of the vortex as the SU(N) Fourier
transform
\begin{align}
\label{emvv}
&e^{-\beta  F_{elec}(\;\vec j\;)}  \\
&\equiv \int 
d\theta_1\int 
d\theta_2 \cdots \int d\theta_{\text{N}-1} \;\;  \chi_{[\vec j]}(\vec \theta)\;
e^{-\beta  F_{mag}(\;\vec \theta\;)}.\nonumber
\end{align}
In (\ref{emvv}), $\chi_{[\vec j]}(\vec \theta)$ is the SU(N) character in the $[\vec j] =(j_1,j_2, \cdots, j_{\text{N}-1})$ 
representation of SU(N). 

The Monte Carlo simulation of $\langle \Sigma^+_{[\vec \theta]} (m,n)\rangle$ in (\ref{dosim})  is problematic because of the presence of infinite Dirac string attached to a vortex  contradicts the periodic boundary conditions imposed on a finite lattice.  On the other hand one can easily compute the 
vortex-anti-vortex correlation functions as shown in Figure \ref{vortexpair}: 
\begin{align}
\begin{aligned}
\langle \Sigma^+_{[\vec \theta]} (m,n) \; \Sigma^-_{[\vec \theta]} (m+R,n)\rangle ~~\equiv ~~e^{-\beta  F(\vec \theta, R)} \\
 = ~~\left\langle   e^{-\beta\sum\limits_{{\sf l}'\in {\cal S}'}\left[\text{tr}(D(\vec \theta)U_p+U_p^\dagger D^\dagger(\vec \theta))-\text{tr}(U_p+U_p^\dagger) \right]}    \right\rangle 
\label{ssdcf}
\end{aligned} 
\end{align}
In (\ref{ssdcf}) ${\cal S}'$ denotes the set of dark links ${\sf l}'$ in Figure \ref{vortexpair} 
and the summation sign  includes only those plaquettes which protrude from  the 
links  ${\sf l}' \in \{{\cal S}'\}$ in the $+$ve time direction.
It will be interesting to study the above free energies and hence the role of SU(N) vortices in the ground state and their magnetic disorder  in the large $R$ limit using
Monte Carlo simulations near  the continuum $\beta \rightarrow \infty$.
\section{Summary and Discussion}

\noindent In this work we have constructed the most general disorder operators for SU(N) lattice gauge theory in $(2+1)$ dimension in the Hamiltonian formulation. Being exactly dual to the Wilson loop operators, these operators create and annihilate $(\text{N}-1)$ types of SU(N) magnetic fluxes. 
The SU(N) order-disorder algebra is simply the canonical commutation relations in the dual formulation, i.e., the commutation relations between the electric scalar potentials and their conjugate magnetic fluxes. 

In the strong coupling limit the disorder \& order operators satisfy: 
 \begin{subequations} 
\begin{eqnarray}
& ~~~\bra{0} \, \,\Sigma^{\pm}_{[\vec \theta]}\,\,\ket{0} &\xrightarrow{~g^2 \rightarrow \infty~}~~1,~~\forall ~[\,\vec \theta\,],  \label{xbb} \\
 & \bra{0} \, \text{Tr}\left({\cal W}^{[\vec j]}_{\cal C}\right) \ket{0}&\xrightarrow{~g^2 \rightarrow \infty~} ~~0,  ~~\, \forall ~[\,\vec j\,]. \label{xaa}
\end{eqnarray}
\end{subequations} 
In  the first limit equation we have used the non-local expression for $\Sigma^{\pm}_{[\vec \theta]}$ in (\ref{disks}).  The strong coupling limits in (\ref{xbb}) \& (\ref{xaa}) show a  complete magnetic disorder atleast in the strong coupling ground state $\ket{0}$.
 The study  of 
 $\langle\Sigma^\pm(\theta)\rangle$ 
 and  the vacuum 
 correlation functions of $\langle \Sigma_\theta^\pm(p)\Sigma_\theta^\mp(p')\rangle$, as $|p-p'|\rightarrow\infty$ for different $[\vec \theta]$ in the weak coupling continuum limit is  required 
 to further probe the relevance of these magnetic disorder operators in the problem of 
 color confinement. These studies across the finite temperature confinement-deconfinement transition will also be useful to understand the magnetic disorder in confining vacuum. 
 We further note that the SU(N) disorder operators are meaningful even in the presence of dynamical matter fields in any SU(N) representation. These canonical transformation techniques can also be generalized to obtain the SU(N)  disorder operator in $(3+1)$ dimension where the dual electric potentials are also the dual gauge fields on the dual links. Thus, like Wilson loop operators ${\cal W}_{[\vec j]}({\cal C})$, the disorder operator $\Sigma_{[\vec \theta]}({\cal C}')$ will also be defined on the closed curves ${{\cal C}'}$ on the dual lattice. The work in these directions is in progress. 
 
\vspace{0.6cm} 

\noindent{\bf Acknowledgements}
\vspace{0.2cm}

\noindent This work is dedicated to the memories of Raja Ji. 
M.M. thanks Ramesh  Anishetty, V. Ravindran and Sayantan Sharma for the invitation to The Institute of Mathematical Sciences (I.M.Sc.), Chennai  where a part of this work was done. M.M. also thanks Ramesh Anishetty for discussions as well as for critical reading of the manuscript.

\appendix
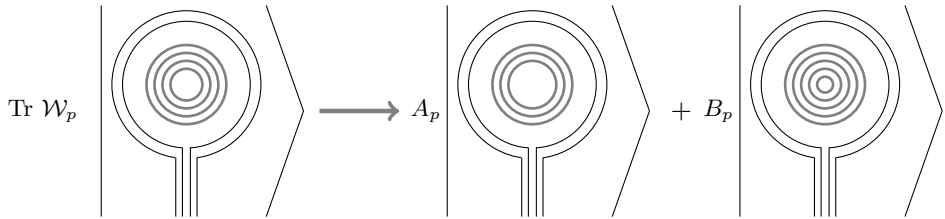
\begin{figure*}
    \centering
    \begin{tikzpicture}[scale=.7]
    \node[] at ( -2.7, -.5) {Tr $\mathcal{W}_p$};
    \draw[] (-1.6, -2.5) --( -1.6, 1.5);
    \draw[] (1.5, -2.5) --(2.2,-.5 )--( 1.5, 1.5);
    \draw[gray,line width=.35 mm] (0,0) circle (.3cm);
    \draw[gray,line width=.35 mm] (0,0) circle (.45cm);
    \draw[gray,line width=.35 mm] (0,0) circle (.6cm);
    \draw[gray,line width=.35 mm] (0,0) circle (.75cm);
    \draw[] (0,0) circle (1.2cm);
    \draw[] (0,0) circle (1.4cm);
    \fill[white] (-.075,-1.15) rectangle (.075,-2.5);
      \fill[white] (-.2,-1.35) rectangle (.2,-2.5);
    \draw[] (-.075, -2.5) --( -.075, -1.19);
    \draw[] (.075, -2.5) --( .075, -1.19);
    \draw[] (-.2, -2.5) --( -.2, -1.38);
    \draw[] (.2, -2.5) --( .2, -1.38);
    \draw[gray, line width=.6mm,->] (2.5,-.5 ) --(4,-.5);
    \begin{scope}[shift={(6.5,0)}]
    \node[] at ( -2, -.5) {$A_p$};
    \draw[] (-1.6, -2.5) --( -1.6, 1.5);
    \draw[] (1.5, -2.5) --(2.2,-.5 )--( 1.5, 1.5);
    \draw[gray,line width=.35 mm] (0,0) circle (.45cm);
    \draw[gray,line width=.35 mm] (0,0) circle (.6cm);
    \draw[gray,line width=.35 mm] (0,0) circle (.75cm);
    \draw[] (0,0) circle (1.2cm);
    \draw[] (0,0) circle (1.4cm);
    \fill[white] (-.075,-1.15) rectangle (.075,-2.5);
        \fill[white] (-.2,-1.35) rectangle (.2,-2.5);
    \draw[] (-.075, -2.5) --( -.075, -1.19);
    \draw[] (.075, -2.5) --( .075, -1.19);
    \draw[] (-.2, -2.5) --( -.2, -1.38);
    \draw[] (.2, -2.5) --( .2, -1.38);
    \end{scope}
    \begin{scope}[shift={(12,0)}]
        \node[scale=1.1] at ( -2.7, -.5) {$+$};
    \node[] at ( -2, -.5) {$B_p$};
    \draw[] (-1.6, -2.5) --( -1.6, 1.5);
    \draw[] (1.5, -2.5) --(2.2,-.5 )--( 1.5, 1.5);
    \draw[gray,line width=.35 mm] (0,0) circle (.15cm);
    \draw[gray,line width=.35 mm] (0,0) circle (.3cm);
    \draw[gray,line width=.35 mm] (0,0) circle (.45cm);
    \draw[gray,line width=.35 mm] (0,0) circle (.6cm);
    \draw[gray,line width=.35 mm] (0,0) circle (.75cm);
    \draw[] (0,0) circle (1.2cm);
    \draw[] (0,0) circle (1.4cm);
    \fill[white] (-.075,-1.15) rectangle (.075,-2.5);
        \fill[white] (-.2,-1.35) rectangle (.2,-2.5);
    \draw[] (-.075, -2.5) --( -.075, -1.19);
    \draw[] (.075, -2.5) --( .075, -1.19);
    \draw[] (-.2, -2.5) --( -.2, -1.38);
    \draw[] (.2, -2.5) --( .2, -1.38);
    \end{scope}
\end{tikzpicture}
    \caption{The action of the Wilson loop on the loop state  $\ket{n=4,l=2,m}$ described in the coupled basis. The circles  in the three figures represent the SU(2) electric  flux circulating in a loop within the plaquette and 
    $2l$ is the number of open flux lines. The action of Tr $\left(\mathcal{W}\right)$ simply  translates $n$ to $n \pm1$ in (\ref{trae}).} 
    \label{fig:tadpole}
\end{figure*}

\section{Electric loop basis} \label{elb} 

\noindent It is easy to construct the loop basis in terms of the dual electric scalar potentials on the plaquette loops
(see Figure \ref{fig:boson}).  In the prepotential representation 
\bea 
{\cal E}^{\rm a}_-(p) \equiv a^\dagger(p) \,\frac{\sigma^{\rm a}}{2}\, a(p),  ~~~~{\cal E}^{\rm a}_+(p) \equiv- b(p)\, \frac{\sigma^{\rm a}}{2}\, b^\dagger(p). 
\label{dmrpp} 
\eea 

Using the facts that the left and the right electric fields  are independent, $[{\cal E}_+^{\rm a}(p),\; {\cal E}_-^{\rm b}(p)] =0$,  and their magnitudes  are equal,  $\sum\limits_{{\rm a}=1}^3 {\cal E}_+^{\rm a}{\cal E}_+^{\rm a}= \sum\limits_{{\rm a}=1}^3{\cal E}_-^{\rm a}{\cal E}_-^{\rm a} \equiv {\cal E}^2$, 
we define the first set of a complete set of commuting operators on every plaquette $p$ as: $\left[{\cal E}^{2},\; {\cal E}_+^{{\rm a}=3},\; {\cal E}_-^{{\rm a}=3}\right]$. 
The SU(2) electric loop decoupled basis on every  plaquette $p$ is 
\begin{align}\label{jmn=phiphi}
\begin{aligned} 
&\ket{j ~m_+ \;m_-} \equiv \ket{j ~m_+} \otimes \ket{j~ m_-}\\ 
& ~~~~~~~~~~~= \phi^j_{m_+}( a^\dagger_1, a^\dagger_2)\ket{0}_a \otimes  \phi^j_{m_-}( b^\dagger_1, b^\dagger_2)\ket{0}_{b},
\end{aligned}  
\end{align}
where we have defined 
\bea 
\phi^j_{m}(a^\dagger_1,a^\dagger_2)=\frac{~~~~~~(a^\dagger_1)^{j+m}}{\sqrt{(j+m)!}}\frac{~~~~~(a^\dagger_2)^{j-m}}{\sqrt{(j-m)!}}. 
\label{2sf} 
\eea
Under SU(2) gauge transformations at the origin  $\Lambda=\Lambda(0,0)$
\begin{align}
\begin{aligned}
\phi^j_{m_+}( a^\dagger_1, a^\dagger_2) ~& \rightarrow ~~ D^{~j}_{{}_{m_+m'_+}}(\Lambda)\; \;\phi^j_{m'_+}( a^\dagger_1, a^\dagger_2) \\
\phi^j_{m_-}( b^\dagger_1, b^\dagger_2)~ & \rightarrow ~\;  D^{~j}_{ m_-m'_-}(\Lambda^\dagger)\; \phi^j_{m'_-}( b^\dagger_1, b^\dagger_2)
\end{aligned} 
\end{align} 
 The electric flux states transform  as 
\begin{equation}
\hspace{-0.1cm}\ket{j ~m_+\;m_-}= \sum_{m'_+, m'_-} D^{~j}_{{m_+m'_+}}(\Lambda)D^{~j}_{ m_-m'_-}(\Lambda^\dagger)\;\ket{j ~m'_+m'_-}.
\label{decb}
\end{equation}

At this stage, it is convenient to work with the coupled basis instead of the decoupled basis (\ref{decb}).  We define the complete set of commuting operators (CSCO) on every plaquette as 
\begin{equation}
\hspace{-0.18cm}\mathcal{E}^2=\mathcal{E}^2_+=\mathcal{E}^2_-,~ L^{\rm a} \equiv \mathcal{E}^{\rm a}_++\mathcal{E}^{\rm a}_-,~ L^{{\rm a}=3} \equiv \mathcal{E}^{{\rm a}=3}_-+\mathcal{E}^{{\rm a}=3}_+.
\end{equation}
The loop coupled basis on every lattice plaquette $\ket{n\;l\;m}$ can be constructed as 
\bea 
\hspace{-0.44cm}\ket{n\;l\;m\;}=&N_{nlm} \;(k_+)^{n-l-1}\,(L_-)^{l-m} (a^\dagger_1)^{l} (b^\dagger_2)^l\;\ket{0,0}
\label{loopbasis1}
\eea
In (\ref{loopbasis1}) $k_+ \equiv a^\dagger \cdot {b}^\dagger \equiv  \sum_{\alpha=1}^2a^\dagger_\alpha {b}^\dagger_\alpha$ 
and 
$$N_{nlm} \equiv \sqrt{\frac{n (l+m)!\, }{(l-m)!\, (l!)^2 (n-l-1)!\,(m+l)!}}.$$ 
The corresponding eigenvalue equations are
\begin{align}
\begin{aligned}
\mathcal{E}^2\;\ket{n\;l\;m}_p~~&=&\;\left(\frac{n^2-1}{4}\right)\;\ket{n\,l\,\,m}_p\\
\vec L^2\;\ket{n\;l\;m}_p ~~&=& l(l+1)~\ket{n\,l\,\,m}_p~~~\\
L^{{\rm a}=3} \;\ket{n\;l\;m}_p~~&=&m~\ket{n\,l\,\,m}_p~~~~~~~~
\end{aligned}
\label{loopbasis} 
\end{align}
In above $ l= 0,1,2, \dots , n-1$ and $m= -l, -l+1, \dots, l$. 
Under gauge transformations  at the origin
$\Lambda = \Lambda(0,0)$, these states have much simpler transformation property
\begin{equation}
\ket{n\;l\;m\;}= \sum_{\bar{m}} D^{~~l}_{m\;\bar{m}} ( \Lambda) \ket{n\; l\; \bar{m}}
\end{equation} 
In other words the  principal $(n)$  and the angular momentum $(l)$ quantum numbers remain 
invariant.

\begin{figure}[h]
    \centering
     \begin{tikzpicture}[scale=1.5]
    \draw[help lines] (0,0) grid (5,4);
   \draw[line width=.5 mm] (0,0)--(2,0)--(2,2)--(4,2)--(4,4)--(0,4)--(0,0);
    \draw[line width=.5mm,->] ( 0,0) --(.5,0);
    \draw[line width=.5mm,->] ( 2,0) --(2,1.5);
    \draw[line width=.5mm,->] ( 2,2) --(2.5,2);
    \draw[line width=.5mm,->] ( 4,2) --(4,2.5);
    \draw[line width=.5mm,->] ( 4,4) --(2.5,4);
    \draw[line width=.5mm,->] ( 0,4) --(0,2.5); 
    \foreach \x in { 0, 1}
    {
    \foreach \y in { 0,1,2,3}
      { 
       \draw[fill] (\x+.5, \y+.5) circle (1.3pt);
      \draw[dotted, thick] ( \x+.25, \y+.1) -- ( \x+.9, \y+.1)-- ( \x+.9, \y+.9)--(\x+.1, \y+.9)-- (\x+.1,\y+.25);
       \draw[thick,->] ( \x+.72,\y+.1)--(\x+.73,\y+.1);
      }
    }
 \foreach \x in { 2, 3}
    {
    \foreach \y in { 2,3}
      { 
       \draw[fill] (\x+.5, \y+.5) circle (1.3pt);
       \draw[dotted, thick] ( \x+.25, \y+.1) -- ( \x+.9, \y+.1)-- ( \x+.9, \y+.9)--(\x+.1, \y+.9)-- (\x+.1,\y+.25); 
       \draw[thick,->] ( \x+.72,\y+.1)--(\x+.73,\y+.1);
      }
    }
\node[scale=.75,below] at ( 3.5,2.45) { $\mathcal{W}(p_1)$};
\node[scale=.75,below] at ( 3.5,3.45) { $\mathcal{W}(p_2)$};
\node[scale=.75,below] at ( 2.5,2.45) { $\mathcal{W}(p_3)$}; 
\node[scale=.75,below] at ( 2.5,3.45) { $\mathcal{W}(p_4)$};
\node[scale=.75,below] at ( 1.5,.45) { $\mathcal{W}(p_5)$};
\node[scale=.75,below] at ( .5,3.45) { $\mathcal{W}(p_{n_c})$};   
\node[scale= 1.05,below] at (2.2,1.65) {$\mathcal{C}$};
    \end{tikzpicture}
    \caption{The Wilson loop ${\cal W(C)}$ of any shape and size can be written as an ordered product of all plaquette operators ${\cal W}(p)$ inside $\cal C$ as in (\ref{wlpl}). These dotted plaquettes inside ${\cal C}$ are illustrated in Figure \ref{fig:boson}.}
    \label{fig:wilsonloop}
\end{figure}
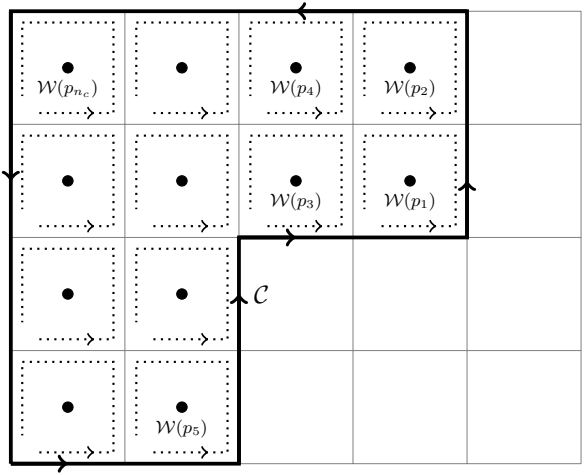

\vspace{0.2cm}
\begin{center}
{\it { The Wilson loops as translation operators }}
\end{center} 
\vspace{0.2cm}
In the loop basis (\ref{loopbasis}), the plaquette operators, ${\cal W}(p)$  which  are unit size  Wilson loop order  operator acts as a translation operator for the electric flux $n$. Using (\ref{plaqhol}) we get     
\begin{align}
\begin{aligned}
\text{Tr} \;\mathcal{W}(p) \;\ket{n\;l\;m}=\;
A_p \; \ket{n-1\; l\; m} +
B_p \;\ket{n+1\; l\; m}.
\end{aligned}
\label{trae}
\end{align}
Here we have ignored the plaquette index p on all the three quantum numbers and 
$$A_p =\frac{ \sqrt{(n-l-1) (n+l)}}{(n-1)}, ~B_p= \frac{ \sqrt{(n+l-1) (n-l)}}{(n+1)}.$$ 
The above translative action of the fundamental plaquette  loop 
operator $\mathcal{W} \equiv \mathcal{W}(p)$ is valid on each plaquette $p$ and we have suppresses 
the plaquette index $p$ on both sides of (\ref{trae}). The  action (\ref{trae}) is illustrated in Figure \ref{fig:tadpole}.  An arbitrary Wilson loop operator ${\cal W}({\cal C})$ can be written in terms of the fundamental plaquette Wilson loop operators 
as 
\bea 
{\cal W}({\cal C}) = \prod_{p \in {\cal C}}\; {\cal W}(p) = {\cal W}(p_1) {\cal W}(p_2) \cdots {\cal W}(p_{n_{\cal C}}). \phantom{xx}
\label{wlpl}
\eea
The above product over plaquettes is taken from the bottom right corner 
as shown in Figure \ref{fig:wilsonloop}. 
Therefore, the end effect of  ${\cal W}({\cal C})$ is to translate the electric fluxes of all plaquette loops inside the closed curve ${\cal C}$ :
\begin{align} 
\begin{aligned}
&{\cal W}({\cal C}) \; \prod_{p\in {\cal C}}\; \ket{n\;l\;m}_p \\&~~= \prod_{p\in {\cal C}}\;\left(A_p \; \ket{n-1\; l\; m}_p+B_p \; \ket{n+1\; l\; m}_p \right). 
\end{aligned} 
\end{align} 
Note that in the SU(N) case the Wilson loop operators will shift all the 
(N-1) Casimir eigenvalues by $\pm 1$. 
\section{ Magnetic loop basis}
\label{mlb}
We now construct the magnetic basis for plaquette flux operators and show that the disorder operator has natural translative action on them.  The group manifold for SU(2) group is $S^3$. We define it on every  plaquette $p$ through  complex doublets $ \vec{z} (p) \equiv (z_1(p),z_2(p)) $ that satisfy the   constraint $ |z_1(p)|^2 + |z_2(p)|^2 =1, \forall ~ p$. A configuration on $S^3$ is 
\bea 
Z(p) = \begin{bmatrix} ~z_1(p) && z_2(p)\\
-z_2^*(p) && z_1^*(p) \\
\end{bmatrix}. 
\label{su2m1}
\eea 
We write eigenvalue equations for the magnetic flux operators as 
\bea \label{wz=zz}
\mathcal{W} _{\alpha\beta}(p) ~|Z(p) \rangle = Z_{\alpha\beta}(p) ~|Z(p)\rangle.
\eea 
Here $Z_{\alpha\beta}(p)$ are the matrix elements of the matrix $Z(p)$ in (\ref{su2m1}). 
These states form a complete orthonormal basis on $S^3$ 
\begin{align}
\begin{aligned} 
&\int_{S^3}\! d\mu(\vec z\;) \;  \ket{Z(p)} \bra{Z(p)}=1, \\
&\braket{Z(p)}{Z'(p)}=\delta (Z(p)-Z'(p)).
\end{aligned} 
\end{align}
The SU(2) group manifold integrations is defined as 
$\int_{SU(2)} d\mu(\vec z\,)
\equiv \frac{1}{16\pi^2}\;\int d^2z_1 d^2z_2 \, \delta(z^*_1z_1+z^*_2z_2-1).$

The magnetic eigenvectors $|Z(p) \rangle$ can be expanded in the complete orthonormal electric basis as
\begin{align}
&\ket{Z( p)}\equiv \;|z_1( p),z_2( p)\rangle \\
&= \sum_{j=0}^\infty \sqrt{(2j+1)}\sum_{m_+, m_-}  D^{~j}_{m_+ \;m_-}(Z (p))~\ket{j\; m_+\; m_-}\nonumber 
\end{align}
The construction of magnetic states can be easily checked by directly applying $\mathcal{W}$ on both sides above equations and realizing that $\mathcal{W}$ acts on the electric field basis as ladder and lowering operators and using the recurrence  relations for the D-functions connecting $D^{~j}_{m_+ \;m_-}$ to  $D^{~~~j\pm \frac{1}{2}}_{m_+\pm\frac{1}{2} \;m_- \pm\frac{1}{2} }$. For SU(N),  $\text{N}\geq 3$, this  approach gets extremely complicated as it requires the recurrence relations 
for the SU(N) Winger D-functions.  We will first write down these states in terms of  SU(2) prepotentials where they take a much simpler form and then  verify the eigenvalues equations (\ref{wz=zz}). Now use equation (\ref{jmn=phiphi})
\begin{align}
\label{zdab}
&\ket{Z( p)}=  \sum_{j=0}^\infty \sqrt{(2j+1)}\\
&\sum_{m_+,\,m_-} D^{~j}_{m_+ ~m_-}(Z ( p))\;\phi^j_{m_+} (a^\dagger_1, a^\dagger_2)\;\phi^j_{m_-} (b^\dagger_1, b^\dagger_2) \ket{0}. \nonumber 
\end{align} 
We call $\phi^j_m(x_1,x_2)$  the SU(2) structure functions. These SU(2) structure functions have the following orthonormal properties: 
\begin{align}
\begin{aligned}
   & \int_{SU(2)} \hspace{-0.4cm} d\mu(\vec z\,) \;
   \phi^{j*}_m (z_1,z_2) \; \phi^j_{m'} (z_1,z_2)=\frac{\delta_{m,m'}}{(2j+1)!} ,\\
   & \sum_{m}\phi^{j*}_m (z_1,z_2) \phi^j_m (w_1,w_2)=\frac{(z^*_1w_1+z^*_2w_2)^{2j}}{(2j)!}.
    \end{aligned}
\end{align}
Further, we can also write SU(2) Wigner D-function in terms of these structure functions as follows:
\begin{equation}
 D^{j}_{{}_{m,n}}\hspace{-0.1cm}(z_1,z_2) =d_j\!\!\int_{{}_{SU(2)}} \hspace{-0.5cm} {d}^2 w_1 {d}^2 w_2 \, \,\phi^{j*}_m (w_1,w_2)\;\phi^{j}_n (z^w_1,z^w_2), ~~~
 \end{equation}
{where} $$\begin{bmatrix}
        z^w_1\\
        z^w_2
    \end{bmatrix} \equiv \begin{bmatrix}
        z_1& z_2\\
        -z_2^* &z^*_1
    \end{bmatrix} \begin{bmatrix}
        w_1\\
        w_2
    \end{bmatrix}, ~~~~d_j \equiv (2j+1)$$
Using properties of structure functions and Wigner D functions, 
$$ \sum\limits_{m_-=-j}^{j} D^j_{{}_{m_+\,m_-}}\hspace{-0.1cm}(z_1,z_2) \; \phi^j_{m_-} (w_1,w_2\;)=\;\phi^j_{m_+} (w^z_1,w^z_2)$$ in (\ref{zdab}), we get
\begin{align}
\ket{Z( p)}=&\sum_{j=0}^\infty \sqrt{(2j+1)}\\
&~\sum_{m_+}\;\phi^j_{m_+} (a^\dagger_1, a^\dagger_2)\;\phi^j_{m_+} ({b}^{z\dagger}_1, {b}^{z\dagger}_2) \ket{0}\nonumber, 
\end{align}
where $$\begin{bmatrix}
        b^{z\dagger}_1\\
        b^{z\dagger}_2
    \end{bmatrix} \equiv \begin{bmatrix}
        z_1& z_2\\
        -z_2^* &z^*_1
    \end{bmatrix} \begin{bmatrix}
        b^\dagger_1\\
        b^\dagger_2
    \end{bmatrix}. $$
Now we can  sum the remaining magnetic index to get
\begin{align}
{\ket{Z ({p})} =\sum_{j=0}^\infty \sqrt{d_j}\,\frac{\left( a^\dagger Z(p){b}^\dagger\right)^{2j}}{(2j)!}\, \ket{0,0}},
\end{align} 
where $d_j=(2j+1)$ is the dimension of  $[j]$ representation. The eigenvalues equation (\ref{wz=zz}) holds at each point of the group manifold. We first  prove it for $Z={I}$ where ${I}$ is the identity element of SU(2) group. First,  we prove that
\begin{equation}
\mathcal{W}^{j=1/2}_{\alpha\beta} ~|{I} \rangle = \delta_{\alpha\beta} ~|{I}\rangle.
\end{equation}
where,
\begin{align}\label{keti}
{\ket{{I}} =\sum_{j=0}^\infty \frac{{(2j+1)^{1/2}}}{(2j)!}( a^\dagger\cdot{b}^\dagger)^{2j}\;\ket{0,0}}
\end{align} 
We have suppressed the plaquette index $p$. Using  prepotential representation (\ref{plaqhol}) for $\mathcal{W}_{\alpha\beta}$ we get
\begin{align*}
\mathcal{W}_{\alpha\beta} \ket{{I}}=& \sum_{j=0}^\infty  \frac{{(2j+1)^{1/2}}}{(2j)!}\left[ \frac{a_\beta  b_\alpha}{(2j)^{1/2}}
- \frac{\tilde{a}^\dagger_\alpha \tilde{b}^\dagger_\beta}{(2j+2)^{1/2}}  \right] \nonumber\\
&~~~~~~~~~~~\frac{1}{\sqrt{(2j+1)}}(a^\dagger \cdot b^\dagger)^{2j}\;\ket{0}\nonumber    
\end{align*}
Now we replace  $ 2j$ by $2j+1$ in the first term and $2j$ by $2j-1$ in the second term of above equation to get 
\begin{align*}
\mathcal{W}_{\alpha\beta} \ket{{I}}=& \sum_{j=0}^\infty  \frac{{(2j+1)^{1/2}}}{(2j)!}\left[ \frac{1}{(2j+1)^2}a_\beta b_\alpha \, (a^\dagger \cdot b^\dagger)^{2j+1}\right.\\
&~~~~~~~~~~~~~~~~~~ \left.- \frac{(2j)}{(2j+1)}a^\dagger_\alpha b^\dagger_\beta (a^\dagger \cdot b^\dagger)^{2j-1}  \right]\ket{0} \nonumber
\end{align*}
we evaluate first term using  the prepotential commutation relations, $ a_\alpha b_\beta ( a^\dagger \cdot b^\dagger)^{2j+1}\,\ket{0}=a_\alpha b_\beta ( \tilde{a}^\dagger \cdot \tilde{b}^\dagger)^{2j+1}\ket{0}=
[(2j+1)^2  \delta _{\alpha\beta}(a^\dagger \cdot b^\dagger)^{2j} + (2j)(2j+1) \tilde{a}^\dagger_\alpha  \tilde{b}^{\dagger}_{\beta} (a^\dagger \cdot b^\dagger)^{2j-1}\;] \ket{0} $ and substitute in above equation to get:
\begin{align*}
\mathcal{W}_{\alpha\beta} \ket{{I}}=& \sum_{j=0}^\infty  \frac{{(2j+1)^{1/2}}}{(2j)!}\delta_{\alpha\beta} (a^\dagger \cdot b^\dagger)^{2j+1}\ket{0}=\delta_{\alpha\beta} \ket{{I}}
\end{align*}
Now we can prove the eigenvalue equation (\ref{wz=zz}), by  considering a transformation of oscillators $b^\dagger_\alpha\rightarrow (Z^\dagger b^\dagger)_\alpha$. Under these transformations $$\mathcal{W}_{\alpha\beta}~ \rightarrow ~ Z^\dagger_{\alpha\gamma} \, \mathcal{W}_{\gamma\beta}, ~~~~~~~~ \ket{{I}}\rightarrow \ket{Z}$$ 
which yields
\bea
Z^\dagger_{\alpha\gamma}\mathcal{W}^{j=1/2}_{\gamma\beta}(p) ~|Z(p) \rangle = ~|Z(p)\rangle.
\eea 
As $Z^\dagger Z ={I}$, we get the eigenvalue equation (\ref{wz=zz}).
 
 The conjugate electric fields act on this basis as differential operators on this plaquette holonomy basis.
\begin{align}
\mathcal{E}^{\rm a}_{+} \ket{Z}=&-\frac{\sigma^{\rm a}_{\alpha\beta}}{2}\, Z_{\gamma\beta}\, \frac{\partial}{\partial Z_{\alpha\gamma}}\, \ket{Z}\\
\mathcal{E}^{\rm a}_{-}\, \ket{Z}=&\;\;\;\;\frac{\sigma^{\rm a}_{\alpha\beta}}{2}\, Z_{\gamma\alpha}\,\frac{\partial}{\partial Z_{\gamma\beta}}\, \;\ket{Z}
\end{align}
For SU(3) these magnetic states are given by
\begin{equation}
\ket{Z}= \sum_{\mathrm{p},\mathrm{q}} \sqrt{d(\mathrm{p},\mathrm{q})}\;\frac{(a^\dagger[1] Z  b^\dagger[1])^\mathrm{p}}{\mathrm{p}!}\;\frac{(a^\dagger[2] Z b^\dagger[2])^\mathrm{q}}{\mathrm{q}!}\ket{0}
\end{equation}
$ d(\mathrm{p}, \mathrm{q})=\frac{1}{2}(\mathrm{p}+1)(\mathrm{q}+1) (\mathrm{p}+\mathrm{q}+2) $ is dimension of $[\mathrm{p},\mathrm{q}]$ representation.
For the general SU(N) case, these magnetic states are given as:
\begin{equation}
\ket{ Z}=\;\sum_{[\vec j]=0}^\infty \sqrt{d(\vec j)} \;\prod_{h=1}^{\text{N}-1} \frac{1}{j_h!}\left( a^
\dagger[h] ~Z ~{b}^\dagger[h]\right)^{2j_h}\; \ket{0}
\end{equation}
Where $d(\vec j)$ is the dimension of the $[\vec j]$ representation and $Z$ represents  (N $\times$ N) SU(N) matrix.

\section{Invisibility of Dirac String}\label{app:invisibility}
\noindent In this Appendix, we explicitly show that  disorder operators in (\ref{disks}) 
creates magnetic flux  only on one plaquette $U_p(m,n)$ located at $(m,n)$. 
They leave all the other plaquettes unaffected. 
The disorder operator involves Kogut-Susskind electric fields $E_-(m,n'\geq n;\hat{1})$, therefore it trivially commutes with all other plaquettes which do not involve 
$U( m,n \geq n'; \hat{1})$. The only relevant plaquette are  $U_p(m,n'\geq n)$. Now we evaluate its action case by case:
\begin{enumerate}
\item First we consider  $n'=n$,  plaquette $U^p\equiv U_p(m,n)$. For convenience we define $U_p(m,n)= U(m-1,n-1;\hat{1})U(m,n-1;\hat {2}) U^\dagger(m-1,n;\hat{1})U^\dagger(m-1,n-1;\hat{2})\equiv U_1U_2U_3^\dagger U_4^\dagger$.  The disorder operator $\Sigma^+_{[\vec{\theta}]}$ will only rotate link $U^\dagger_3$ around the axis  $\vec{\Theta}^{\rm a}_0= R^{\rm ab}(U_1U_2)\vec{\theta}^{\;\rm b} $: 
\begin{align*}
\Sigma^+_{[\vec{\theta}]}U^p_{\alpha\beta}\Sigma^{+\dagger}_{[\vec{\theta}]}
=& \left[U_1U_2\left(\Sigma^+_{[\vec{\theta}]}U^\dagger_3\Sigma^{+\dagger}_{[\vec{\theta}]} \right)\;U^\dagger_4\right]_{\alpha\beta}\\
=& \left[U_1U_2 \;D(\vec{\Theta}_0)U^\dagger_3\; U^\dagger_4\right]_{\alpha\beta}\\
=& \left[U_1U_2\;D(U_2^\dagger U_1^\dagger\,\vec{\theta}\,U_1U_2)\; U^\dagger_3\; U^\dagger_4\right]_{\alpha\beta}
\end{align*}
Now we use a property of Wigner D-matrices namely $D(U_2^\dagger U_1^\dagger\;\vec{\theta}\,U_1U_2)=U_2^\dagger U_1^\dagger D(\vec{\theta}\;)U_1U_2$ to get 
\begin{align*}
\Sigma^+_{[\vec{\theta}]}\;U^p_{\alpha\beta}\;\Sigma^{+\dagger}_{[\vec{\theta}] }=\left[D (\vec{\theta}) U^p\right]_{\alpha\beta}.
\end{align*}
Therefore, the disorder operators $\Sigma^+_{[\vec{\theta}]}(m,n)$ create magnetic flux at plaquette $U_p(m,n)$.
\item For $n'>n$, consider plaquette $U^p\equiv U_p(m,n')$ For convenience we define $U_p(m,n')= U(m-1,n'-1;\hat{1})U(m,n'-1;\hat {2}) U^\dagger(m-1,n';\hat{1})U^\dagger(m-1,n'-1;\hat{2})\equiv U_1U_2U_3^\dagger U_4^\dagger$.  The disorder operator $\Sigma^+_{[\vec{\theta}]}$ will rotate two horizontal links $U_1$ and $U_3^\dagger$  around the axes $\vec{\Theta}_{n'}$ and $\vec{\Theta}_{n'+1}$ respectively. Due to equation (\ref{theta}), these two axes are  related as   $\vec{\Theta}^{\rm a}_{n'+1}= R^{\rm ab}(U_2)\vec{\Theta}^{\;\rm b}_{n'}$:
\begin{align*}
\Sigma^+_{[\vec{\theta}]}U^p_{\alpha\beta}\Sigma^{+\dagger}_{[\vec{\theta}]}
&=\left[\left(\Sigma^+_{[\vec{\theta}]}U_1\Sigma^{+\dagger}_{[\vec{\theta}]}\right) U_2\left(\Sigma^+_{[\vec{\theta}]}\,U^\dagger_3\,\Sigma^{+\dagger}_{[\vec{\theta}]}\right)\, 
U^\dagger_4\right]_{\alpha\beta}\\
&=\;\left[U_1D^\dagger(\vec{\Theta}_{n'})U_2 D(\vec{\Theta}_{n'+1})U^\dagger_3U^\dagger_4\right]_{\alpha\beta}\\
&=\;\left[U_1D^\dagger(\vec{\Theta}_{n'})U_2 D(U_2^\dagger \,\vec{\Theta}_{n'}\,U_2 )U^\dagger_3U^\dagger_4\right]_{\alpha\beta}\\
&=\;\left[U_1D^\dagger(\vec{\Theta}_{n'})U_2 U_2^\dagger  D(\,\vec{\Theta}_{n'} )\,U_2U^\dagger_3U^\dagger_4\right]_{\alpha\beta}\\
&= U^p_{\alpha\beta}
\end{align*}
Therefore, the disorder operators $\Sigma^+_{[\vec{\theta}]}(m,n)$ leave all plaquette $U_p(m,n'>n)$ unaffected.
\end{enumerate}
Thus we have shown that  the disorder operator rotates a plaquette by a Wigner D matrix and hence creates a  SU(N) magnetic vortex there.
Now we will show that the shape of the Dirac string which is a vertical line $\mathcal{S}$ in Figure \ref{dstring} can be deformed by gauge transformations and is therefore unphysical.
 Deformation of Dirac string by unit plaquette at site $(m,n')$ can be achieved by replacing $ E^{\rm a}_+(m,n';\hat{1})$ by $- (E^{\rm a}_-(m,n';\hat{1})+E^{\rm a}_+(m,n';\hat{2})+E^{\rm a}_-(m,n';\hat{2}))$ in the equation (\ref{disks}). Deformed string $\tilde{\cal S}$ is shown in Figure \ref{dstring}. Applying similar replacements we can change the shape of the Dirac string arbitrarily with a fixed endpoint.

\bibliographystyle{unsrt}

\begin{thebibliography}{10}

\bibitem{Note1}
This result follows from the SU(N) Young tableau in the $[\protect \vec j]$
  representation with total $L$ fundamental boxes. If each of them is rotated
  by the center element $z$ then we get $D^{[\protect \vec j]}(z) = (z)^{L}
  \protect \tmspace +\thickmuskip {.2777em}{\protect \cal I} = (z)^{\eta
  [\protect \vec j]} \protect \tmspace +\thickmuskip {.2777em}{\protect \cal
  I}$ as $z^\protect \text {N}=1$. Here ${\protect \cal I} $ is the identity
  matrix in the $[\protect \vec j]$ representation of SU(N).

\bibitem{Note2}
The convention chosen for loop (string) electric fields is that $ \protect
  \mathcal {E}^{\protect \rm a}_{-}(\protect \vec n) ~(\protect \mathbb
  {E}^{\protect \rm a}_{-}( \protect \vec n))$ and $\protect \mathcal
  {E}^{\protect \rm a}_{+}(\protect \vec n)~(\protect \mathbb {E}^{\protect \rm
  a}_{+}(\protect \vec n))$ are located at the initial, end points of the flux
  loop (string) respectively.

\bibitem{Note3}
We have used the relation $n^{\protect \rm a}(p)\protect \mathcal {E}^{\protect
  \rm a}_-(p) =- n^{\protect \rm a}(p) R^{\protect \rm ab}(\protect \mathcal
  {W}^\dagger (p))\protect \mathcal {E}^{\protect \rm b}_+(p)$ and \protect
  \vspace {-0.11cm} \begin {align*} \protect \phantom {XX} n^{\protect \rm
  a}(p) R^{\protect \rm ab}&(\protect \mathcal {W}^\dagger (p))\\&=\protect
  \text {Tr} (\sigma ^{\protect \rm a} \protect \mathcal {W}(p))\protect
  \tmspace +\thickmuskip {.2777em}\protect \frac {1}{2}\protect \text
  {Tr}(\sigma ^{\protect \rm a} \protect \mathcal {W}^\dagger (p) \sigma
  ^{\protect \rm b}\protect \mathcal {W}(p))\\ &=\protect \tmspace
  +\thickmuskip {.2777em}\protect \frac {1}{2}(\sigma ^{\protect \rm a}_{\alpha
  \beta }\sigma ^{\protect \rm a}_{\gamma \delta }) \sigma ^{\protect \rm
  b}_{\eta \rho } \protect \mathcal {W}_{\beta \alpha }(p)\protect \mathcal
  {W}^\dagger _{\delta \eta }(p)\protect \mathcal {W}_{\rho \gamma }(p)\\
  &=\protect \tmspace +\thickmuskip {.2777em}\protect \frac {1}{2}(2\delta
  _{\alpha \delta }\delta _{\beta \gamma } -\delta _{\alpha \beta }\delta
  _{\gamma \delta })\sigma ^{\protect \rm b}_{\eta \rho }\protect \mathcal
  {W}_{\beta \alpha }(p)\protect \mathcal {W}^\dagger _{\delta \eta
  }(p)\protect \mathcal {W}_{\rho \gamma }(p)\\ &= \sigma ^{\protect \rm
  b}_{\eta \rho }\protect \mathcal {W}_{\rho \eta }(p)=n^{\protect \rm
  b}(p)\label {nrn} \end {align*}.

\bibitem{Note4}
In defining ${\protect \cal E}_-^a(p)$ we have used the fact that like
  (${\sigma }^a$), their transpose with a negative sign ($-\protect \tilde
  {\sigma }^a$) also satisfies the same SU(2) Lie algebra.

\bibitem{Note5}
The two definitions for $\protect \vec {n}_{[1]}(p)$ and $\protect \vec
  {n}_{[2]}(p)$ in (\ref {Su3axes1}) and (\ref {Su3axes2}) respectively are
  easily generalizable to SU(N) case discussed in the next section.

\bibitem{Note6}
We have used the following two identities\\ $1.~ (f^{{\protect \rm a}{\protect
  \rm b}{\protect \rm e}}d^{{\protect \rm d}{\protect \rm c}{\protect \rm
  e}}+f^{{\protect \rm a}{\protect \rm c}{\protect \rm e}}d^{{\protect \rm
  b}{\protect \rm d}{\protect \rm e}}+f^{{\protect \rm a}{\protect \rm
  d}{\protect \rm e}}d^{{\protect \rm b}{\protect \rm c}{\protect \rm e}})=0$\\
  $2.~ (d^{{\protect \rm a}{\protect \rm b}{\protect \rm e}}d^{{\protect \rm
  d}{\protect \rm c}{\protect \rm e}}+d^{{\protect \rm a}{\protect \rm
  c}{\protect \rm e}}d^{{\protect \rm b}{\protect \rm d}{\protect \rm
  e}}+d^{{\protect \rm a}{\protect \rm d}{\protect \rm e}}d^{{\protect \rm
  b}{\protect \rm c}{\protect \rm e}})=\protect \tfrac {1}{3}(\delta
  ^{{\protect \rm a}{\protect \rm b}}\delta ^{{\protect \rm c}{\protect \rm
  d}}+\delta ^{{\protect \rm a}{\protect \rm c}}\delta ^{{\protect \rm
  b}{\protect \rm d}}+\delta ^{{\protect \rm b}{\protect \rm c}}\delta
  ^{{\protect \rm a}{\protect \rm d}})$.

\bibitem{Note7}
We are ignoring the SU(3) multiplicity problem here as the aim in this work is
  to construct the SU(3) magnetic eigenstates and not to worry about SU(3)
  multiplicities. One can trivially replace all SU(3) prepotentials by SU(3)
  irreducible prepotentials \cite {manpp} at the end without changing any
  results of this section. The same strategy will be adapted in the next SU(N)
  section to keep the discussion simple.

\bibitem{Note8}
Advantage of this representation is that it has the following property: \begin
  {align*} &Z( \omega _1, \omega _2)Z( \theta _1, \theta _2)\\ &~~~~= e^{\left
  (i (\omega _1\protect \hat {n}^{\protect \rm a}_{[1]} + \omega _2\protect
  \hat {n}^{\protect \rm a}_{[2]}) \protect \frac {\lambda ^{\protect \rm
  a}}{2}\right )}e^{\left (i (\theta _2\protect \hat {n}^{\protect \rm b}_{[1]}
  + \theta _2\protect \hat {n}^{\protect \rm b}_{[2]}) \protect \frac {\lambda
  ^{\protect \rm b}}{2}\right )}\\ &~~~~~~~~ \because e^X e^Y= e^{X+Y+\protect
  \frac {1}{2}[X,Y]+...},~~~~ [\lambda ^{\protect \rm a},\lambda ^{\protect \rm
  b}]=2if^{{\protect \rm a}{\protect \rm b}{\protect \rm c}}\lambda ^{\protect
  \rm c},\\ &~~~~~~~~~~~~~~~~~~~~~~~~~~~~~f^{{\protect \rm a b c}} \protect
  \hat {n}^{\protect \rm b}_{[h]}\protect \hat {n}^{\protect \rm
  c}_{[h']}=0,~~~~h,h'=1,2\\ &~~~~~= e^{\left (i ((\omega _1+\theta _1)\protect
  \hat {n}^{\protect \rm a}_{[1]} + (\omega _1+\theta _2)\protect \hat
  {n}^{\protect \rm a}_{[2]}) \protect \frac {\lambda ^{\protect \rm
  a}}{2}\right )}\\ &~~~~~=Z(\omega _1+\theta _1,~\omega _2+\theta _2) \end
  {align*} Which we will use to show the translation of two gauge invariant
  angles through the action of the disorder operator.

\bibitem{Note9}
We have used the property: $(d^{{\protect \rm ab e}}d^{{\protect \rm cd
  e}}+d^{{\protect \rm ac e}}d^{{\protect \rm bd e}}+d^{{\protect \rm
  ade}}d^{{\protect \rm bc e}})=\protect \tfrac {1}{3}(\delta ^{{\protect \rm a
  b}}\delta ^{{\protect \rm c d}}+\delta ^{{\protect \rm a c}}\delta
  ^{{\protect \rm b d}}+\delta ^{{\protect \rm b c}}\delta ^{{\protect \rm a
  d}})$ for SU(3). It can be similarly generalized to SU(N) with $(\protect
  \text {N}-1)$ d structure functions.

\bibitem{Note10}
Here we have redefined the parallel transport ${\protect \sf S}(m,n;n')\equiv
  {\protect \sf T}^\dagger (m-1,n)S(m,n;n')$ and magnetic axis $\protect \vec
  {n}^{\protect \rm a}_{[1]}(m,n)\equiv R^{\protect \rm ab}({\protect \sf T}
  (m-1,n))\protect \vec {n}^{\protect \rm b}_{[1]}(p)= \protect \text {Tr} (
  \Lambda ^{\protect \rm a}( U_p(m,n) +U^\dagger _p(m,n)))$. The advantage of
  using new parallel transports ${\protect \sf S}(m,n;n')$ is that they are not
  connected to the origin and therefore more appropriate for the original
  Kogut-Susskind formulation.

\end{thebibliography}


\begin{thebibliography}{99}
\bibitem{kada} L. P. Kadanoff and H. Ceva, ``Determination of an Operator Algebra for the Two-Dimensional Ising Model", \href{https://journals.aps.org/prb/abstract/10.1103/PhysRevB.3.3918}{Phys. Rev. B \textbf{3}, 3918 (1971)};
\bibitem{kramers} H. A. Kramers and G. H. Wannier, ``Statistics
of the two-dimensional ferromagnet. part I", \href{https://journals.aps.org/pr/abstract/10.1103/PhysRev.60.252}{Phys. Rev. \textbf{60}, 252 (1941)}.

\bibitem{frad} E. Fradkin, L. Susskind, ``Order and disorder in gauge systems and magnets", \href{https://journals.aps.org/prd/abstract/10.1103/PhysRevD.17.2637}{Phys. Rev. D \textbf{17}, 2637(1978)}; E. Fradkin, ``Disorder Operators and Their Descendants
", \href{https://link.springer.com/article/10.1007/s10955-017-1737-7}{J. Statist. Phys. \textbf{167}, 427 (2017)}; \href{https://arxiv.org/abs/1610.05780}{arXiv:1610.05780}; Fradkin, Eduardo, ``Field theories of condensed matter physics", \href{https://inspirehep.net/literature/327355}{Cambridge University Press (2013)};  
P. R. Braga, M. S. Guimaraes, M. A. Paganelly, ``Multivalued fields and monopole operators in topological superconductors", \href{https://www.sciencedirect.com/science/article/pii/S0003491620301792}{Ann. Phys. \textbf{419} 168245 (2020)}.

\bibitem{wegner} D. Horn, M. Weinstein, and S. Yankielowicz ``Hamiltonian approach to Z(N) lattice gauge theories", \href{https://journals.aps.org/prd/abstract/10.1103/PhysRevD.19.3715}{Phys. Rev. D 19, 3715}, ,  F. J. Wegner, ``Duality in Generalized Ising Models and Phase Transitions without Local Order Parameters", \href{https://pubs.aip.org/aip/jmp/article-abstract/12/10/2259/465334/Duality-in-Generalized-Ising-Models-and-Phase?redirectedFrom=fulltext}{J. Math. Phys. \textbf{12}, 2259-2272 (1971)}.


\bibitem{guth} Akira Ukawa, Paul Windey, and Alan H. Guth, ``Dual variables for lattice gauge theories and the phase structure of Z(N) systems", \href{https://journals.aps.org/prd/abstract/10.1103/PhysRevD.21.1013}{Phys. Rev. D \textbf{21}, 1013 (1980)}; I. G. Halliday, A. Schwimmer, ``The phase structure of SU(N)/Z (N) lattice gauge theories", \href{https://www.sciencedirect.com/science/article/pii/0370269381900551}{Physics Letters B \textbf{101}(5) 327-331 (1981)}. 

\bibitem{kogut51} J. B. Kogut,  An introduction to lattice gauge theory and spin systems", \href{https://journals.aps.org/rmp/pdf/10.1103/RevModPhys.51.659}{Rev. Mod. Phys \textbf{51}(4) 659 (1979)}. 

\bibitem{man1} S. Mandelstam, Berkeley preprint (1978). S. Mandelstam, ``Feynman Rules for Electromagnetic and Yang-Mills Fields from the Gauge-Independent Field-Theoretic Formalism",  \href{https://journals.aps.org/pr/abstract/10.1103/PhysRev.175.1580}{Phys. Rev. \textbf{175}, 1580 (1968)}; S. Mandelstam, ``Charge-monopole duality and the phases of non-Abelian gauge theories", \href{https://journals.aps.org/prd/abstract/10.1103/PhysRevD.19.2391}{Phys. Rev. D \textbf{19}, 2391 (1979)}; S. Mandelstam, ``General introduction to confinement" 
\href{https://www.sciencedirect.com/science/article/pii/0370157380900836}{Phys. Rep. \textbf{67}, No. 1 (1980) 1-199}; S. Mandelstam, ``Vortices and quark confinement in non-abelian gauge theories", \href{https://www.sciencedirect.com/science/article/pii/037026937590221X}{Physics Letters B. 53: 476-478}; S. Mandelstam,  ``II. Vortices and quark confinement in non-Abelian gauge theories",  \href{https://www.sciencedirect.com/science/article/pii/0370157376900430}{Phys. Rep. \textbf{23}: 245-249}. 
\bibitem{hooft} 	G. 't Hooft, ``On the phase transition towards permanent quark confinement", 
\href{https://www.sciencedirect.com/science/article/pii/0550321378901530}{Nucl. Phys. B\textbf{138} 1-25 (1978)}; G. 't Hooft, ``A property of electric and magnetic flux in non-Abelian gauge theories", \href{https://www.sciencedirect.com/science/article/pii/0550321379905959}{Nucl. Phys. B \textbf{153} 141-160 (1979)}. 

\bibitem{mack} G. Mack, V. B. Petkova, ``Comparison of lattice gauge theories with gauge groups $Z_2$ and SU (2)", \href{https://www.sciencedirect.com/science/article/pii/0003491679903464}{Ann. Phys. \textbf{123}(2) 442-467 (1979)};   G. Mack, E. Pietarinen, ``Monopoles, Vortices, and Confinement", \href{https://www.sciencedirect.com/science/article/pii/0550321382903819?via%3Dihub}{Nucl. Phys. B \textbf{205} 141 (1982)}; G. Mack, ``Quark Confinement in Lattice Gauge Theories", \href{https://inspirehep.net/literature/157259}{Acta Phys. Austriaca Suppl. \textbf{22} 509-529 (1980)};
  G. Mack, V.B. Petkova, ``$Z_2$ Monopoles in the Standard SU(2) Lattice Gauge Theory Model", \href{https://inspirehep.net/literature/140990}{Z. Phys. C \textbf{12} 177 (1982)}.

\bibitem{tomb} E. Tomboulis, ``'t Hooft loop in SU(2) lattice gauge theories", \href{https://journals.aps.org/prd/abstract/10.1103/PhysRevD.23.2371}{Phys. Rev. D, \textbf{23} 10 2371 (1981)}; Laurence G. Yaffe, ``Confinement in SU(N) lattice gauge theories", 
\href{https://journals.aps.org/prd/abstract/10.1103/PhysRevD.21.1574}{Phys. Rev. D \textbf{21}, 1574 (1980)}.

\bibitem {dis_nonabelian} 
 T. Kovacs, E. Tomboulis, ``Computation of the vortex free energy in SU(2) gauge theory", \href{https://journals.aps.org/prl/abstract/10.1103/PhysRevLett.85.704}{Phys.
Rev. Lett. \textbf{85}, 704 (2000)}, \href{https://arxiv.org/abs/hep-lat/0002004}{arXiv:hep-lat/0002004};  Philippe de Forcrand, David Noth, ``Precision lattice calculation of SU(2) 't Hooft loops", \href{https://journals.aps.org/prd/pdf/10.1103/PhysRevD.72.114501}{Phys. Rev. D \textbf{72}, 114501 (2005)}; H. Reinhardt and D. Epple, ``The 't Hooft loop in the Hamiltonian approach to Yang-Mills theory in Coulomb gauge", \href{https://journals.aps.org/prd/pdf/10.1103/PhysRevD.76.065015}{Phys. Rev. D \textbf{76}, 065015 (2007)};  T. Kovacs, E. Tomboulis, ``Vortices and confinement",   \href{https://inspirehep.net/literature/512435}{NATO Sci. Ser. C \textbf{553} 315-326 (2000)};
T. Kovacs, E. Tomboulis, ``Vortex structure of the vacuum and confinement", \href{https://inspirehep.net/literature/535827}{Nucl. Phys. B Proc. Suppl. \textbf{94} 518-521 (2001)}:
J. Greensite, ``The Confinement problem in lattice gauge theory", \href{https://inspirehep.net/literature/612137}{Prog. Part. Nucl. Phys. \textbf{51} 1(2003)}; 
\bibitem{magnetic_disorder}  
K. Langfeld, H. Reinhardt and O. Tennert,`` Confinement and scaling of the vortex vacuum of SU(2) lattice gauge theory ", \href{https://www.sciencedirect.com/science/article/pii/S0370269397014354}{Phys. Lett. B \textbf{419} 316 (1998)}; L. Del Debbio, M. Faber, J. Greensite, and S. Olejnik, `` Center dominance and $Z_2$ vortices in SU(2) lattice gauge theory",
\href{https://journals.aps.org/prd/abstract/10.1103/PhysRevD.55.2298}{Phys. Rev. D \textbf{55} 2298 (1997)}


 \bibitem{reinhardt} H. Reinhardt, ``On 't Hooft's loop operator", \href{https://www.sciencedirect.com/science/article/pii/S0370269303001990}{Phys. Lett. B \textbf{557}(3-4), 317-323 (2003)}; Takuya Shimazaki and Arata Yamamoto, ``'t Hooft surface in lattice gauge theory",
\href{https://journals.aps.org/prd/abstract/10.1103/PhysRevD.102.034517}{Phys. Rev. D \textbf{102}, 034517 (2020)}.

\bibitem{jordanwigner} P. Jordan, . Der Zusammenhang der symmetrischen und linearen Gruppen und das Mehrkorper problem. Zeitschrift fur Physik, \textbf{94}(7-8), 531-535 (1935), P. Hasenfratz,
A puzzling combination: Disorder $\times$ order, Physics Letters B, Volume 85, Issue 4, 1979,
Pages 338-342.

\bibitem{mansre}  Manu Mathur and T. P. Sreeraj, ``Lattice gauge theories and spin models", \href{https://journals.aps.org/prd/abstract/10.1103/PhysRevD.94.085029}{Phys. Rev. D \textbf{94}, 085029 (2016)}.Manu Mathur and T. P. Sreeraj, ``Canonical transformations and loop formulation of SU (N) lattice gauge theories", \href{https://journals.aps.org/prd/abstract/10.1103/PhysRevD.92.125018}{Phys. Rev. D \textbf{92}, 125018(2015)}.

\bibitem{mmar} Manu Mathur, Atul Rathor, ``Exact Duality and Local Dynamics in SU(N ) Lattice Gauge Theory", \href{https://journals.aps.org/prd/abstract/10.1103/PhysRevD.107.074504}{Phys. Rev. D \textbf{107}, 074504 (2023)}. 


\bibitem{manpp} M. Mathur, ``Harmonic oscillator prepotentials in SU (2) lattice gauge theory", \href{https://iopscience.iop.org/article/10.1088/0305-4470/38/46/008}{J. Phys. A: Math. Gen. \textbf{38}(46) 10015 (2005)}; R. Anishetty, M. Mathur and I. Raychowdhury, ``Prepotential formulation of SU (3) lattice gauge theory", \href{https://iopscience.iop.org/article/10.1088/1751-8113/43/3/035403}{J. Phys. A:
Math. Theor. \textbf{43}(3) 035403 (2009)}: M. Mathur, I. Raychowdhury and R. Anishetty, "SU (N) irreducible Schwinger Bosons", \href{https://pubs.aip.org/aip/jmp/article/51/9/093504/896813/SU-N-irreducible-Schwinger-Bosons}{Journal of mathematical physics \textbf{51}(9) 093504(2010)}.

\bibitem{helleretc} Urs M. Heller, ``String tension in (2+1)-dimensional compact lattice QED: Weak and strong-coupling results; a variational calculation" \href{https://journals.aps.org/prd/abstract/10.1103/PhysRevD.23.2357}{Phys. Rev. D \textbf{23}, 2357 (1981)}; S. D. Drell, H.R. Quinn, B. Svetitsky and M. Weinstein,
``{QED} on a Lattice: A Hamiltonian Variational Approach to the Physics of the Weak Coupling Region", \href{https://journals.aps.org/prd/abstract/10.1103/PhysRevD.19.619}{Phys. Rev. D \textbf{19} 619 (1979)}; N.E. Ligterink, N.R. Walet, R.F. Bishop, ``Toward a Many-Body Treatment of Hamiltonian Lattice SU(N) Gauge Theory",  \href{https://www.sciencedirect.com/science/article/pii/S0003491600960706}{Annals Phys. 284 (2000) 215-262};  John B. Bronzan, ``Explicit Hamiltonian for SU(2) lattice gauge theory", \href{https://journals.aps.org/prd/abstract/10.1103/PhysRevD.31.2020}{Phys. Rev. D 31 (1985), 2020};  John B. Bronzan, ``Analytic approach to weak-coupling SU(2) Hamiltonian lattice gauge theory", \href{https://journals.aps.org/prd/pdf/10.1103/PhysRevD.37.1621}{Phys. Rev. D 37 (1988), 1621}.  




\bibitem{kogsus} J. B. Kogut and L. Susskind, ``Hamiltonian formulation of Wilson's lattice gauge theories", \href{https://journals.aps.org/prd/abstract/10.1103/PhysRevD.11.395}{Phys. Rev. D \textbf{11}, 395 (1975)}.




\bibitem{mallesh1997algebra} K. S. Mallesh, N. Mukunda,  ``The algebra and geometry of SU(3) matrices", \href{http://repository.ias.ac.in/78455/}{Pramana -Journal of Physics, \textbf{49}(4) 371-383 (1997) ISSN 0304-4289}. 

\bibitem{georgi2000lie} H. Georgi, `` Lie algebras in particle physics: from Isospin To Unified Theories", \href{https://library.oapen.org/bitstream/handle/20.500.12657/50876/9780429967764.pdf?sequence=11}{Westveiw press (1999)}.
\end{thebibliography}
\end{document}